\documentclass[]{aastex701}
\usepackage{placeins}
\usepackage{xeCJK} 
\usepackage{amsmath}

\begin{document}


\title{The Local Tremaine-Weinberg Method for Galactic Pattern Speed: Theory and its Application to IllustrisTNG}

\author[0000-0003-1022-4617]{Hangci Du (杜航慈)}
\affiliation{National Astronomical Observatories \\ Chinese Academy of Sciences \\ Beijing 100101, China}
\affiliation{School of Astronomy and Space Science\\University of Chinese Academy of Sciences\\Beijing 100049, China}
\affiliation{Innovation Academy for Microsatellites\\ Chinese Academy of Sciences\\ Shanghai 201304, China}
\email[show]{hcdu@bao.ac.cn}

\author{Yougang Wang (王有刚)}
\affiliation{National Astronomical Observatories \\ Chinese Academy of Sciences \\ Beijing 100101, China}
\affiliation{School of Astronomy and Space Science\\University of Chinese Academy of Sciences\\Beijing 100049, China}
\email[show]{wangyg@bao.ac.cn}

\author{Junqiang Ge (葛均强)}
\affiliation{National Astronomical Observatories \\ Chinese Academy of Sciences \\ Beijing 100101, China}
\affiliation{School of Astronomy and Space Science\\University of Chinese Academy of Sciences\\Beijing 100049, China}
\email[show]{jqge@nao.cas.cn}

\author{Rui Guo (郭锐)}
\affiliation{National Astronomical Observatories \\ Chinese Academy of Sciences \\ Beijing 100101, China}
\affiliation{School of Astronomy and Space Science\\University of Chinese Academy of Sciences\\Beijing 100049, China}
\email[show]{}


\begin{abstract}
The Tremaine-Weinberg (TW) method and its variations provide the most direct means to measure the pattern speeds of galactic bars. We establish a unifying framework by deriving an integral form of the continuity equation over an arbitrary closed loop. This naturally defines a local pattern speed for any chosen region in a galactic disk (including bars and spirals). We demonstrate that this intuitive formalism recovers all standard variants of the TW method as special cases corresponding to specific choices of the integration loop. 
In this paper, we validate this framework and demonstrate its diagnostic power. By applying it to a diverse set of test cases from the TNG50 simulation, including face-on prototype barred galaxies and highly constrained Mock Milky Way standard configurations, we show that this formalism accurately recovers both constant global pattern speeds and radially varying profiles. Rather than relying on rigid geometric approximations, our method naturally differentiates coherent solid-body rotators (bars) from spirals. Our results validate that this unified integral framework provides a robust, geometrically flexible, and practically extensible tool for decoding complex dynamics of galactic structures.
\end{abstract}


\section{Introduction} \label{sec:intro}

Non-axisymmetric structures, such as stellar bars and spiral arms, are ubiquitous features in disk galaxies and are believed to be primary drivers of their secular evolution. Observational surveys indicate that a substantial fraction, perhaps as many as two-thirds, of nearby disk galaxies host a bar structure \citep{1999ASPC..187...72K, 2000AJ....119..536E}. These features are not merely static patterns; they are dynamically active, rotating entities that profoundly influence the distribution and kinematics of both stellar and gaseous components within their host galaxies. They act as powerful engines for redistributing angular momentum, driving large-scale gas inflows towards the galactic center, triggering bursts of star formation, and shaping the overall morphological and chemical evolution of the disk \citep{1993RPPh...56..173S, 2003MNRAS.341.1179A, 2014RvMP...86....1S}.

For barred galaxies, the pattern speed, $\Omega_p$, is the key parameter for linking the visible structure of these galaxies to their internal dynamics and long-term evolution. This parameter describes the angular velocity at which the bar or spiral pattern rotates. The value of $\Omega_p$ dictates the locations of the principal dynamical resonances within the galactic disk, such as the Inner and Outer Lindblad Resonances (ILR, OLR) and the corotation resonance (CR), where stars and gas orbit at the same angular speed as the pattern itself. These resonances are loci of significant dynamical influence, mediating the exchange of energy and angular momentum between the pattern and the disk materials \citep{2008gady.book.....B}. Consequently, an accurate measurement of $\Omega_p$ is fundamental for constraining the orbital structure of the disk, understanding the mechanisms of gas fueling to active galactic nuclei, and testing models of galaxy formation and evolution. For instance, the ratio of the corotation radius to the bar's length, $\mathcal{R} \equiv R_{\rm CR} / R_{\rm bar}$, is a key diagnostic for classifying ``fast" ($\mathcal{R} \lesssim 1.4$) or ``slow" ($\mathcal{R} > 1.4$) bars. This ratio is sensitive to the angular momentum exchange between the bar and the dark matter halo, making its measurement a critical observational constraint on the properties and content of dark matter in galaxies \citep{2000ApJ...543..704D}.

Direct methods aim to measure $\Omega_p$ from the first principle. The most prominent and widely adopted method is the one proposed by \citet{1984ApJ...282L...5T}, commonly known as the Tremaine-Weinberg (TW) method. 
Under the assumptions of a stable, rigidly rotating pattern and a conserved tracer population (e.g., stars or a specific gas phase), the TW method relates the pattern speed to luminosity-weighted moments of the tracer's surface brightness and line-of-sight velocity field. As it relies on a fundamental conservation law rather than a detailed dynamical model, it is considered one of the most robust and direct observational techniques available. The recent advent of integral field unit (IFU) spectroscopy has ushered in a new era for such measurements, enabling the application of the TW method to large, statistically significant samples of galaxies from surveys like CALIFA \citep{2015A&A...576A.102A} and MaNGA \citep{2023MNRAS.521.1775G}. These studies have provided unprecedented insights into the demographics of bar pattern speeds, revealing correlations with host galaxy properties and suggesting that bars may universally form as fast rotators and subsequently slow down through various evolutionary processes \citep{2020A&A...641A.111C, 2023MNRAS.521.1775G}. 

The conceptual simplicity of the TW method has inspired the development of a diverse diverse array of related techniques, each adapted for a specific observational scenario or scientific goal. These include the classic formulation for external galaxies using long-slit spectra \citep{1995MNRAS.274..933M}; methods tailored for our internal perspective within the Milky Way, which must contend with different geometric projections and observational challenges \citep{2002MNRAS.334..355D, 2019MNRAS.488.4552S}; and ``radial" TW methods designed to probe potential variations of the pattern speed with galactocentric radius, particularly relevant for complex structures like multi-component bars or spiral arms \citep{2008ApJ...676..899M, 2016ApJ...826....2S}. While powerful in their respective domains, these methods have been developed and derived largely in isolation. 

This diverse landscape calls for a unifying perspective. While recent years have seen the development of sophisticated numerical algorithms—capable of measuring the ``ground truth" pattern speed with very high precision in N-body simulation and providing an invaluable benchmark fo testing observational techniques  \citep[e.g.,][]{2023MNRAS.518.2712D,2024A&A...691A.122H, 2024A&A...692A.159S, 2025ApJ...978...37A}. A parallel effort has been made to develop local or regional methods for analyzing observational data, particularly for systems with multiple patterns (e.g., \citealt{2023A&A...673A..36P}). The latter, however, typically reply on the differential  form of the continuity equation and still operate without a general, unifying physical picture.     

In this context, the goal of this paper is not to develop another competing numerical algorithm to achieve maximum precision, but to construct a transparent framework that fosters further physical understanding. We introduce a framework based on a general integral form of the continuity equation. To this end, our primary objectives are threefold: (1) to introduce the concept of a ``local pattern speed" defined via a generalized integral form of the continuity equation, demonstrating that all aforementioned TW-like methods emerge as special cases of this single formalism; (2) to leverage this unified view to clarify the implicit approximations and assumptions inherent in each method; and (3) to apply this framework to propose a theoretically geometrically exact formulation for measuring radially varying pattern speeds from observational data. In doing so, we aim to provide a clear, physically intuitive, and extensible foundation for understanding, comparing, and developing methods to probe the dynamics of non-axisymmetric structures in galaxies. 

This paper is structured as follows. In Section 2, we develop our local pattern speed framework. We build the geometric and physical intuition from the continuity equation, present its generalization for arbitrary two- and three-dimensional geometries, and delineate the fundamental assumptions. We also establish this framework as a unifying parent theory for existing Tremaine-Weinberg methods, with detailed mathematical derivations provided in the Appendices. Section 3 applies this methodology to a diverse sample of disk galaxies from the TNG50 cosmological simulation, demonstrating the framework's diagnostic power across typical bars, transient spiral waves, and complex multi-component systems. Finally, Section 4 synthesizes our discussion, summarizes the core conclusions, and outlines future applications.

\section{The Local Pattern Speed Framework}
\label{sec:framework}

In this section, we develop our framework for measuring pattern speeds. We start from establishing the fundamental geometric and physical intuition behind our approach, which is rooted in the continuity equation. We then formalize this intuition into a general integral expression that defines the ``local pattern speed'' over an arbitrary region in the galactic disk. Finally, we discuss the key assumptions underlying the framework and its interpretation in the context of multiple patterns.

\subsection{Intuitive Derivation: The Annular Sector}
\label{sec:intuition}

We begin by establishing a direct geometric interpretation of the pattern speed, complementing the abstract differential continuity equation in favor of a tangible ``mass balance" argument. This approach allows us to define the pattern speed $\Omega_{\rm p}$ based purely on the observable fluxes of tracers across a specified boundary.

\begin{figure}[ht!]
    \centering
    \includegraphics[width=0.7\textwidth]{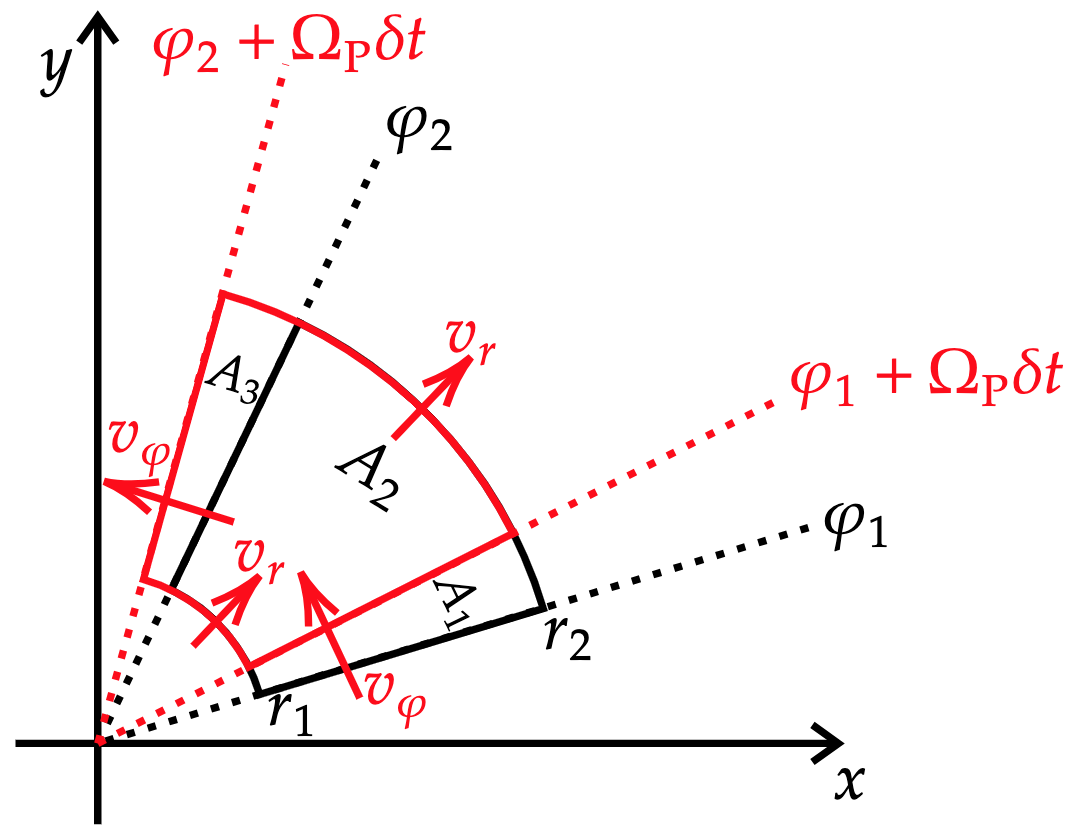} 
    \caption{Direct geometric interpretation of the local pattern speed. A coherent galactic pattern rotates around the origin with angular velocity $\Omega_{\rm p}$. We verify the mass balance within a small annular sector (bounded by red or black lines). During a time interval $\delta t$, the pattern rotates from the black sector ($A_1 \cup A_2$) to the red sector ($A_2 \cup A_3$). Provided that the tracer particles are neither created nor destroyed, the mass difference between the non-overlapping regions $A_1$ and $A_3$ must obtain a precise balance with the net mass flux crossing the sector boundaries.}
    \label{fig:local-geo}
\end{figure}

Consider a tracer population (e.g., stars or gas) with surface density $\Sigma(r, \varphi)$ and velocity field $\mathbf{v} = v_r \hat{\mathbf{e}}_r + v_\varphi \hat{\mathbf{e}}_\varphi$. Let us focus on a specific local region defined by an annular sector bounded by radii $r \in [r_1, r_2]$ and azimuths $\varphi \in [\varphi_1, \varphi_2]$, as illustrated in Figure~\ref{fig:local-geo}.

Imagine a galactic pattern---such as a spiral arm or a bar segment---rotating rigidly around the galactic center with a constant angular speed $\Omega_{\rm p}$. At an initial time $t$, this pattern occupies the region delineated by the black boundaries in Figure~\ref{fig:local-geo}, comprising sub-regions $A_1$ and $A_2$. After a short time interval $\delta t$, the pattern has rotated by an angle $\delta \phi = \Omega_{\rm p} \delta t$, now occupying the red region comprising $A_2$ and $A_3$.

The fundamental principle governing this system is the conservation of mass. If we assume that the tracer population is conserved (i.e., source and sink terms are negligible), then any change in the mass configuration must be accounted for by the physical motion of stars or gas. Specifically, the mass difference between the ``new" pattern region ($A_3$) and the ``old" pattern region ($A_1$) must be equal to the net mass flux flowing across the boundaries of the fixed sector during the interval $\delta t$.

Mathematically, as $\delta t \to 0$, we can express the mass change due to the pattern's rotation (the ``sweeping" effect) as the integral of the surface density over the swept area. Simultanously, the net flux is the integral of the velocity components normal to the boundaries. This balance can be written as a system of relations:
\begin{equation}
    \left \{ 
    \begin{aligned} 
        \delta M_{\rm pattern} & \approx \delta \int \Sigma \, dS = \Omega_{\rm p} \delta t \left[ \int_{r_1}^{r_2} \Sigma(r, \varphi_2) \, r \, dr - \int_{r_1}^{r_2} \Sigma(r, \varphi_1) \, r \, dr \right] \\ 
        & = \Omega_{\rm p} \delta t \left. \left( \int_{r_1}^{r_2}\Sigma \, r \, dr \right) \right|_{\varphi_1}^{\varphi_2}, \\
        \delta M_{\rm flux} & \approx - \delta t \oint_{\partial A} (\Sigma \mathbf{v}) \cdot \hat{\mathbf{n}} \, dl \\
        & = \delta t \left[ \left.\left( \int_{r_1}^{r_2} \Sigma v_\varphi \, dr \right)\right|_{\varphi_1}^{\varphi_2} + \left.\left( \int_{\varphi_1}^{\varphi_2} \Sigma v_r r \, d\varphi \right)\right|_{r_1}^{r_2} \right].
    \end{aligned} 
    \right.
    \label{eq:mass_change_sector}
\end{equation}
Here, the first equation represents the mass displaced by the pattern's rotation, while the second equation sums the physical influx and outflux of tracers across the azimuthal and radial boundaries. The notation $\left. f \right|_a^b$ denotes $f(b) - f(a)$.

Equating the pattern-induced mass displacement to the physical mass flux ($\delta M_{\rm pattern} = \delta M_{\rm flux}$) allows us to solve directly for $\Omega_{\rm p}$. This yields a closed-form expression for the pattern speed measured within this specific annular sector:
\begin{equation}
    \Omega_{\rm p} = \frac
    {\left.\left(\int_{r_1}^{r_2} \Sigma v_\varphi dr\right)\right|_{\varphi_1}^{\varphi_2} + \left.\left(\int_{\varphi_1}^{\varphi_2} \Sigma v_r rd\varphi\right)\right|_{r_1}^{r_2}}
    {\left.\left(\int_{r_1}^{r_2}\Sigma rdr\right)\right|_{\varphi_2}^{\varphi_1}}.
    \label{eq:local_pattern_speed}
\end{equation}
Equation~\eqref{eq:local_pattern_speed} is significant because it defines the pattern speed locally. Unlike the classic Tremaine-Weinberg method, which requires integration over the entire extent of the galaxy along a slit, this formulation allows for the determination of $\Omega_{\rm p}$ using only data from the boundaries of a finite ``local" region. This intuitive picture forms the conceptual basis for the generalized integral formulation we develop next, which will allow us to re-derive all TW-like methods from a single, unified starting point and explore new, optimized integration geometries. 

\subsection{Generalization to Arbitrary Geometries}
\label{sec:general_integral}

\begin{figure}[ht!]
\centering
\includegraphics[width=0.6\textwidth]{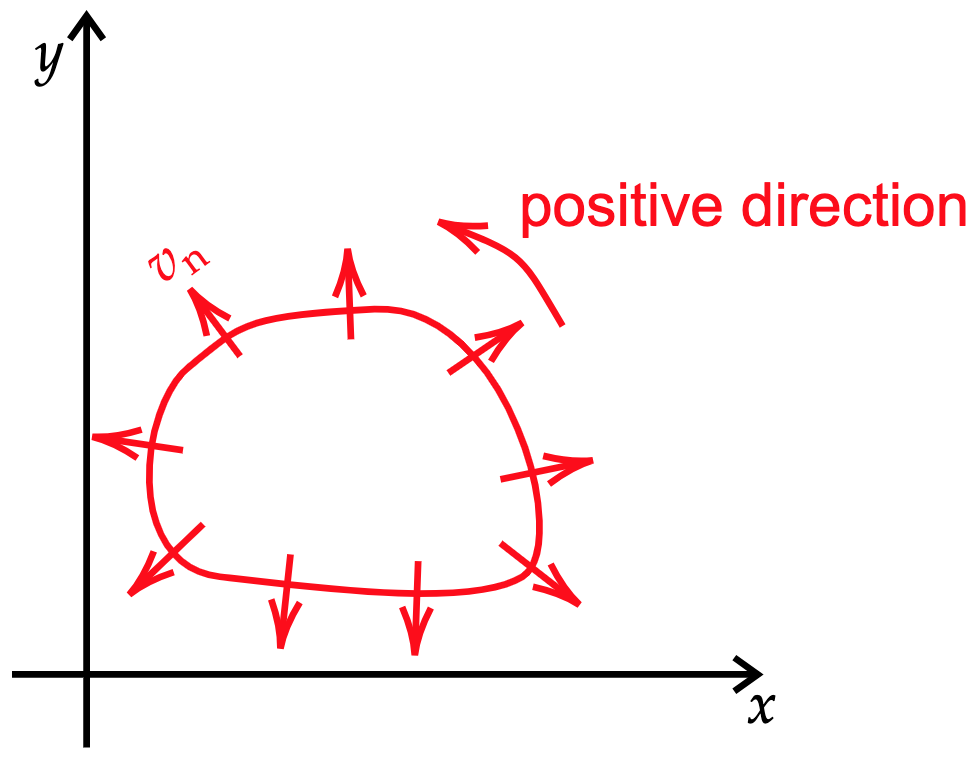}
\caption{
A unified view of the local pattern speed measurement based on the continuity equation, generalized from Figure~\ref{fig:local-geo}. The local pattern speed $\Omega_{\rm p}$ is defined by line integrals around an arbitrary closed path $\partial S$ enclosing an area $S$. For the pattern to appear stationary in a co-rotating frame, the tracer mass must be conserved. This requires that the net physical mass flux out of the region, given by the numerator of Equation~(\ref{eq:general_loop_integral}), $\oint \Sigma (\mathbf{v} \cdot \hat{\mathbf{n}}) dl$, must be perfectly balanced by the effective flux induced by the rigid rotation of the pattern, given by the denominator, $\Omega_{\rm p} \oint \Sigma ((\hat{\mathbf{z}} \times \mathbf{r}) \cdot \hat{\mathbf{n}}) dl$. Here, $\hat{\mathbf{n}}$ is the outward normal vector to the boundary element $d\mathbf{l}$. This formulation is universal, independent of the loop's shape, and directly connects the pattern speed to the observable kinematics ($\mathbf{v}$) and density ($\Sigma$) along any chosen boundary.
}
\label{fig:unified-geo}
\end{figure}

While Equation~\eqref{eq:local_pattern_speed} provides a clear physical intuition, the annular sector geometry is restrictive. Real-world observational data or complex simulation outputs may benefit from integration paths that conform to specific morphological features (e.g., following the curvature of a spiral arm) or data coverage footprints. We therefore seek to generalize this concept to an arbitrary closed loop $\partial S$ enclosing an area $S$, as depicted in Figure~\ref{fig:unified-geo}. Based on intuitive generalization or vector calculus, we can obtain the following general integral form:

\begin{equation}
    \Omega_{\rm p} = -\frac{\oint_{\partial S} \Sigma v_n \, dl}{\oint_{\partial S} \Sigma \mathbf{r} \cdot d\mathbf{l}}.
    \label{eq:general_loop_integral}
\end{equation}

For details of the derivation, the reader is referred to Appendix~\ref{sec:general_derivation}

\subsection{Extension to Three Dimensions}

With the advent of high-precision astrometric surveys such as \textit{Gaia}, full 6D phase-space information (position and velocity) is available for large samples of stars. It is therefore highly advantageous to extend this formalism to three dimensions. The logic remains unchanged, but we now consider a volume $V$ bounded by a closed surface $\partial V$ with volume density $\rho$.

Starting from the 3D continuity equation and applying the 3D divergence theorem, Equation~\eqref{eq:continuity_rotating_expanded} generalizes directly. The pattern speed $\Omega_{\rm p}$ (still representing rotation about the $z$-axis) is given by:
\begin{equation}
    \Omega_{\rm p} = \frac{\int_V \nabla \cdot (\rho \mathbf{v}) \, dV}{\int_V \nabla \cdot (\rho \hat{\mathbf{z}} \times \mathbf{r}) \, dV} 
    = \frac{\oint_{\partial V} \rho v_n \, dS}{\oint_{\partial V} (\hat{\mathbf{z}} \times \rho \mathbf{r}) \cdot d\mathbf{S}},
    \label{eq:local_omega_p_3d}
\end{equation}
where $d\mathbf{S} = \hat{\mathbf{n}} dS$ is the vector surface element.

Equation~\eqref{eq:local_omega_p_3d} is particularly powerful for analyzing the Milky Way. Unlike 2D methods that require projecting the galaxy onto a plane or assuming a cylindrically consistent density, this 3D formulation allows for the calculation of the pattern speed using a volume of tracers—for instance, a sphere centered on the Sun or a box in the Galactic disk—accounting for vertical motions and density gradients explicitly. This provides a robust theoretical foundation for local pattern speed determination in the era of precision 3D astrometry.

\subsection{On the Assumptions}
\label{sec:assumptions}

The general integral formulation presented in Section~\ref{sec:general_integral} offers a powerful and flexible tool for measuring pattern speeds, unifying various methods under a single physical principle. However, the accuracy and interpretability of any measured $\Omega_{\rm p}$ rest on a set of fundamental assumptions. The choice of tracer population and integration region must be made with a clear understanding of their respective domains of validity. In this section, we delineate these core assumptions and discuss their implications.

\begin{enumerate}
    \item \textbf{Tracer Conservation.} Central to our framework is the continuity equation, which presupposes that the chosen tracer is a conserved quantity on the dynamical timescales of interest. That is, particles of the tracer population are neither created nor destroyed within the integration volume. This assumption is generally well-satisfied for mature stellar populations (e.g., red giant branch stars), which have stellar evolution timescales far exceeding the galaxy's rotation period. However, for gaseous tracers such as \ion{H}{1} or H$\alpha$, this assumption is more tenuous. Gas can be converted into stars, particularly within the high-density regions of spiral arms, or be affected by feedback processes, introducing effective source and sink terms that are not accounted for. Indeed, hydrodynamical simulations demonstrate that when applied to gaseous components, the Tremaine-Weinberg method often fails to recover the true pattern speed precisely because these non-gravitational effects violate the simple continuity equation \citep{2023MNRAS.524.3437B}. Therefore, care must be taken when interpreting pattern speeds derived from gas kinematics.

    \item \textbf{A Stationary and Rigidly Rotating Pattern.} Our derivation assumes the existence of a single, well-defined pattern that rotates rigidly and is stationary in its co-rotating frame. This implies that the pattern speed, $\Omega_{\rm p}$, is constant in both space and time. 
    
    The assumption of a time-constant $\Omega_{\rm p}$ requires careful consideration in a cosmological context. While an isolated, stable bar may maintain a nearly constant pattern speed, numerical simulations consistently show that bar pattern speeds evolve over Gyr timescales. A classic mechanism involves the transfer of angular momentum from the bar to the dark matter halo via dynamical friction, causing $\Omega_{\rm p}$ to decrease over time \citep{1984ApJ...282L...5T, 2008gady.book.....B}. This secular slowdown can be counteracted by a gas-rich disk, which can absorb angular momentum or fuel star formation, helping to ``lock'' the bar's rotation and maintain a more stable pattern speed \citep{2023ApJ...953..173B}. Recent results from the TNG50 cosmological simulation paint a detailed picture of this evolution. On average, bar pattern speeds have decreased by roughly a factor of two from redshift $z=1$ to the present day \citep{2024A&A...691A.122H}. This evolution is strongly mass-dependent: bars in massive galaxies tend to slow down rapidly as AGN feedback expels cold gas, leaving dynamical friction as the dominant mechanism, while bars in lower-mass galaxies with more gas can maintain a more constant $\Omega_{\rm p}$ for longer periods \citep{2024A&A...692A.159S}. Nonetheless, for the purpose of analyzing a single observational snapshot, this long-term evolution is not a primary concern. The assumption of a stationary pattern remains valid as long as the timescale for changes in $\Omega_{\rm p}$ is much longer than the dynamical timescale of the galaxy, a condition that is generally met.

    A more immediate challenge is the assumption of a single, rigid pattern. Galaxies are complex systems that can host multiple, co-existing patterns with different speeds, such as a primary bar and a distinct set of spiral arms. N-body simulations frequently show that local spiral arms can rotate with a different pattern speed than the main bar, significantly perturbing the local kinematics. In such a scenario, applying a method that assumes a single $\Omega_{\rm p}$ to a region containing multiple patterns will yield a luminosity-weighted average of their speeds. As we will discuss further in Section \ref{sec:radial_tw}, our framework can, in principle, be adapted to disentangle these components, but the simplest application (Eq.~\ref{eq:general_loop_integral}) implicitly assumes a single dominant pattern within the integration area $S$.

    \item \textbf{Planar Geometry.} Our 2D formulation (Eq.~\ref{eq:general_loop_integral}) inherently assumes that all motion is confined to a flat plane. This is an excellent approximation for the inner regions of most disk galaxies, where the bar and main spiral structures reside. However, in the outer disk, phenomena such as galactic warps and flaring become significant, causing stars and gas to deviate substantially from a single plane. Applying the 2D method in such regions can introduce systematic errors. For analyses extending into these warped and flared regions, or for detailed studies of the Milky Way where full 3D positional and velocity information is available, our 3D formulation (Eq.~\ref{eq:local_omega_p_3d}) provides the more appropriate and robust theoretical foundation.
\end{enumerate}

In summary, while our integral framework is mathematically general, its practical application requires a careful selection of both the tracer population and the integration region, guided by the scientific question at hand and the physical conditions of the target galaxy. Acknowledging these assumptions is the first step toward mitigating their potential effects and obtaining robust measurements of galactic pattern speeds.

\subsection{Interpretation with Multiple Pattern Speeds}
\label{sec:multiple_patterns}

\begin{figure}[ht!]
\centering
\includegraphics[width=0.7\textwidth]{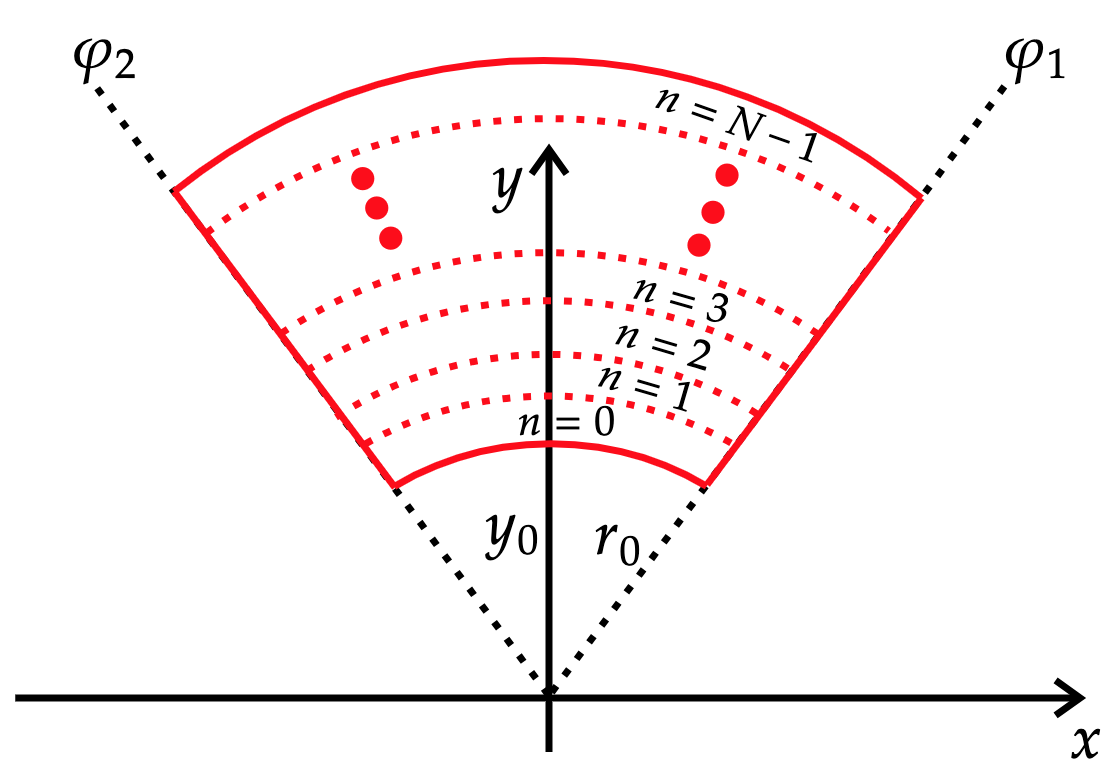} 
\caption{Schematic illustrating the application of the local pattern speed framework to a region harboring multiple, radially varying pattern speeds. The annular sector, bounded by the solid red lines, is conceptually subdivided into a series of thinner annuli (separated by dotted red lines), with each sub-annulus $n$ assumed to host a pattern rotating with a distinct local speed $\Omega_{\rm p}(r_n)$. As derived in Section~\ref{sec:multiple_patterns}, applying our integral formulation (Eq.~\ref{eq:general_loop_integral}) across the entire outer boundary yields a single measured value, $\langle \Omega_{\rm p} \rangle$. This value is not a simple average, but a weighted mean of the individual speeds $\Omega_{\rm p}(r_n)$, where the weights are determined by the strength of the non-axisymmetric density signature within each sub-annulus. This highlights how the choice of integration path explicitly determines which dynamical components are being measured.}
\label{fig:multiple-patterns}
\end{figure}

The framework developed thus far rests on the idealization of a single, rigidly rotating pattern. However, real galaxies are dynamically complex. It is well established from both theoretical work and hydrodynamical simulations that galaxies can simultaneously host multiple non-axisymmetric components---such as a bar and a set of spiral arms---that rotate with different pattern speeds \citep[e.g.,][]{1988MNRAS.231P..25S, 2006ApJ...639..868S}. Modern cosmological simulations consistently show that spiral structures are often transient features with pattern speeds that vary with radius, and which can exist independently of a central bar structure \citep{2011MNRAS.410.1637S, 2026ApJ...997..363Q}. For example, local spiral arms near the Sun may rotate at a different speed than the Milky Way's bar, significantly influencing local kinematics. This raises a critical question: what does our integral formulation measure when applied to a region where $\Omega_{\rm p}$ is not constant?

Here, we demonstrate that when the integration loop $\partial S$ encloses regions with varying pattern speeds (see Figure~\ref{fig:multiple-patterns}), the resulting measurement, which we denote $\langle \Omega_{\rm p} \rangle$, is a weighted average of the constituent speeds. Crucially, the weighting is not simply proportional to the tracer mass but is determined by the local strength of the pattern's non-axisymmetric density signature:

\begin{equation}
    \langle \Omega_{\rm p} \rangle = \frac{\sum_{n=0}^{N-1} \Omega_{\rm p}(r_n) \cdot w_n}{\sum_{n=0}^{N-1} w_n},
    \label{eq:weighted_average_omega}
\end{equation}
where the weight $w_n$ for each sub-annulus is
\begin{equation}
    w_n = \int_{r_n}^{r_{n+1}} \left[ \Sigma(r, \varphi_1) - \Sigma(r, \varphi_2) \right] r \, dr.
    \label{eq:weight_definition}
\end{equation}

For details of the derivation, the reader is referred to Appendix~\ref{sec:multiple_derivation}.

\subsection{Theoretical Unification}
\label{sec:theoretical_unification}

The general integral formulation for the local pattern speed, derived in Section~\ref{sec:framework} and summarized in Equation~(\ref{eq:general_loop_integral}), provides a powerful and versatile tool. Its true utility lies not only in enabling new measurement strategies but also in establishing a unified theoretical foundation for all existing methods based on the continuity equation. 

In Appendix~\ref{sec:tw_special_cases}, we provide rigorous derivations demonstrating how our formalism unifies the following techniques:
\begin{itemize}
    \item \textbf{The Classic Tremaine-Weinberg (TW) Method:} We show that the traditional slit-based method for external galaxies reduces to our general equation when integrating over a specific, infinitely large loop (Appendix~\ref{sec:classic_tw}).
    \item \textbf{Milky Way Bar and Spiral Arm Methods:} We demonstrate that modern techniques utilizing proper motions for the Galactic bar (e.g., viewing configurations through semi-infinite cylinders) and line-of-sight velocities for local spiral arms (spherical volumes) are direct 3D geometrical manifestations of our framework (Appendices~\ref{sec:mw_bar} and \ref{sec:mw_spirals}).
    \item \textbf{Radially Varying Pattern Speeds:} By analyzing the flux balance, our framework elucidates a subtle geometric approximation inherent in the traditional continuous ``radial" TW method. Building upon this insight, we construct a geometrically exact matrix formulation for recovering radially varying pattern speeds limits (Appendices~\ref{sec:radial_tw} and \ref{sec:refined_radial_tw}).
\end{itemize}

This unified perspective elegantly clarifies the explicit and implicit geometric weightings, as well as the fundamental assumptions, underlying each established technique.

\section{The Diversity of Pattern Speed Profiles in TNG50}
\label{sec:diversity_profile}

After validating the method on an individual textbook case, we turn to the broader TNG50 population. We apply the method to a representative selection of massive disk galaxies ($M_* > 10^{10} M_\odot$) from TNG50-1 at $z=0$, chosen to illustrate the diversity of dynamical states naturally occurring in a cosmological context. 

TNG50-1 is the highest-resolution realization of the IllustrisTNG project \citep{2019MNRAS.490.3196P, 2019MNRAS.490.3234N}. TNG50 constitutes a cosmological gravo-magnetohydrodynamical simulation evolved using the moving-mesh code \textsc{Arepo}. It traces galaxy formation and evolution within a comoving volume of $(51.7 \, \text{Mpc})^3$, achieving a baryonic mass resolution of $8.5 \times 10^4 M_{\odot}$ and a spatial resolution defined by a gravitational softening length of 288\,pc for stars at $z \le 1$ \citep{2019MNRAS.490.3196P, 2024MNRAS.535.1721P}. The simulation's adaptive gas refinement can reach scales as small as 72\,pc. This unprecedented combination of large-scale volume and high spatial fidelity is critical for resolving the internal stellar kinematics required for our local pattern speed analysis. Indeed, TNG50 has proven capable of producing realistic analogs of the Milky Way and Andromeda galaxies \citep{2024MNRAS.535.1721P}.

The structural realism of TNG galaxies has been extensively validated against observations. For instance, mock observations processed through radiative transfer codes demonstrate that TNG100 galaxies exhibit non-parametric morphological indices (e.g., Gini–$M_{20}$, concentration, asymmetry) consistent with Pan-STARRS data within $1\sigma$ uncertainties \citep{2019MNRAS.483.4140R}. Furthermore, the kinematic scaling relations (e.g., Faber-Jackson), central density slopes, and structural parameters of early-type galaxies in the simulation show remarkable agreement with observational constraints \citep{2025MNRAS.539.2855D}.

Specifically regarding barred galaxies, TNG50 provides a robust statistical sample for dynamical studies. We utilize the catalog of barred galaxies identified by \citet{2022MNRAS.512.5339R}, which serves as the basis for our sample selection. While comparisons with MaNGA observations suggest that bars in TNG50 may be statistically shorter than observed counterparts, their kinematic properties remain physically consistent \citep{2022ApJ...940...61F}. Recent studies have also utilized this dataset to explore the secular evolution of bar pattern speeds \citep{2024A&A...691A.122H, 2024A&A...692A.159S}. Complementing these global analyses, this section aims to demonstrate how a radially resolved perspective—enabled by the local pattern speed framework—unveils kinematic details and diversity often obscured in global measurements.

\subsection{Typical Bars and the Non-detection of Ultrafast Bars}
\label{sec:typical_bars}

\begin{figure*}[ht!]
\centering
\fig{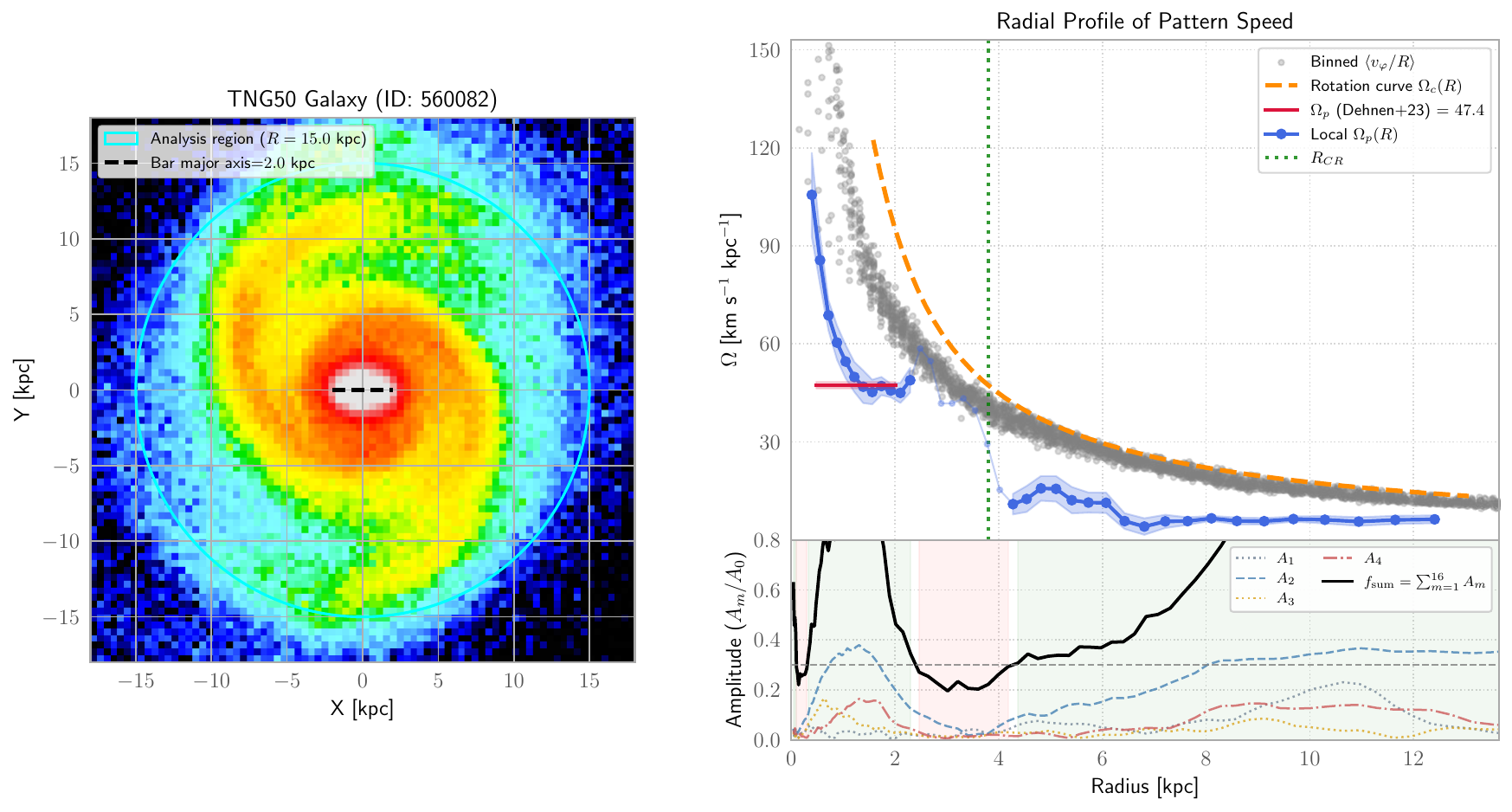}{0.7\textwidth}{(a) Grand-design Spiral (ID 560082)}

\fig{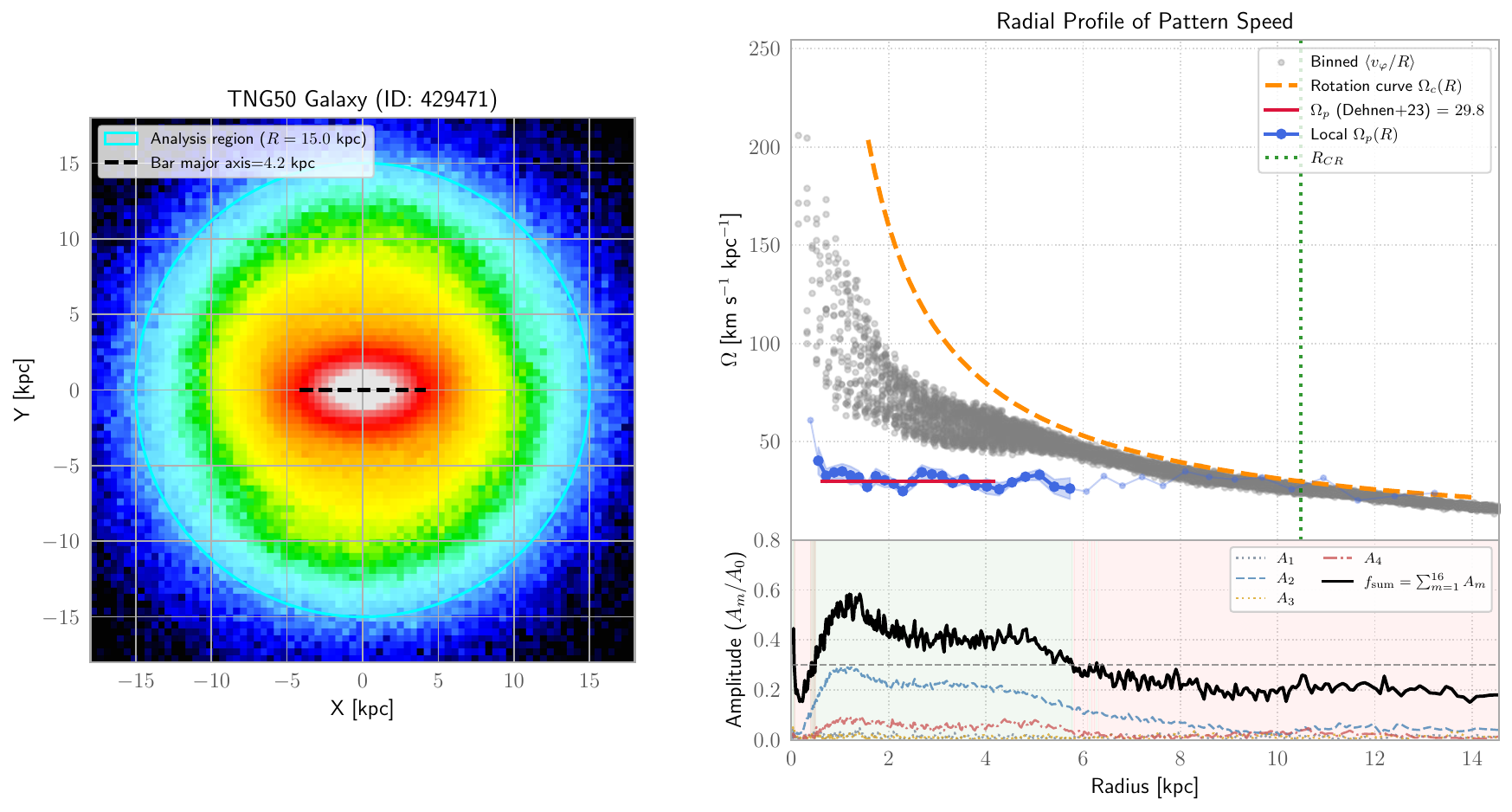}{0.7\textwidth}{(b) Isolated Bar (ID 429471)}

\fig{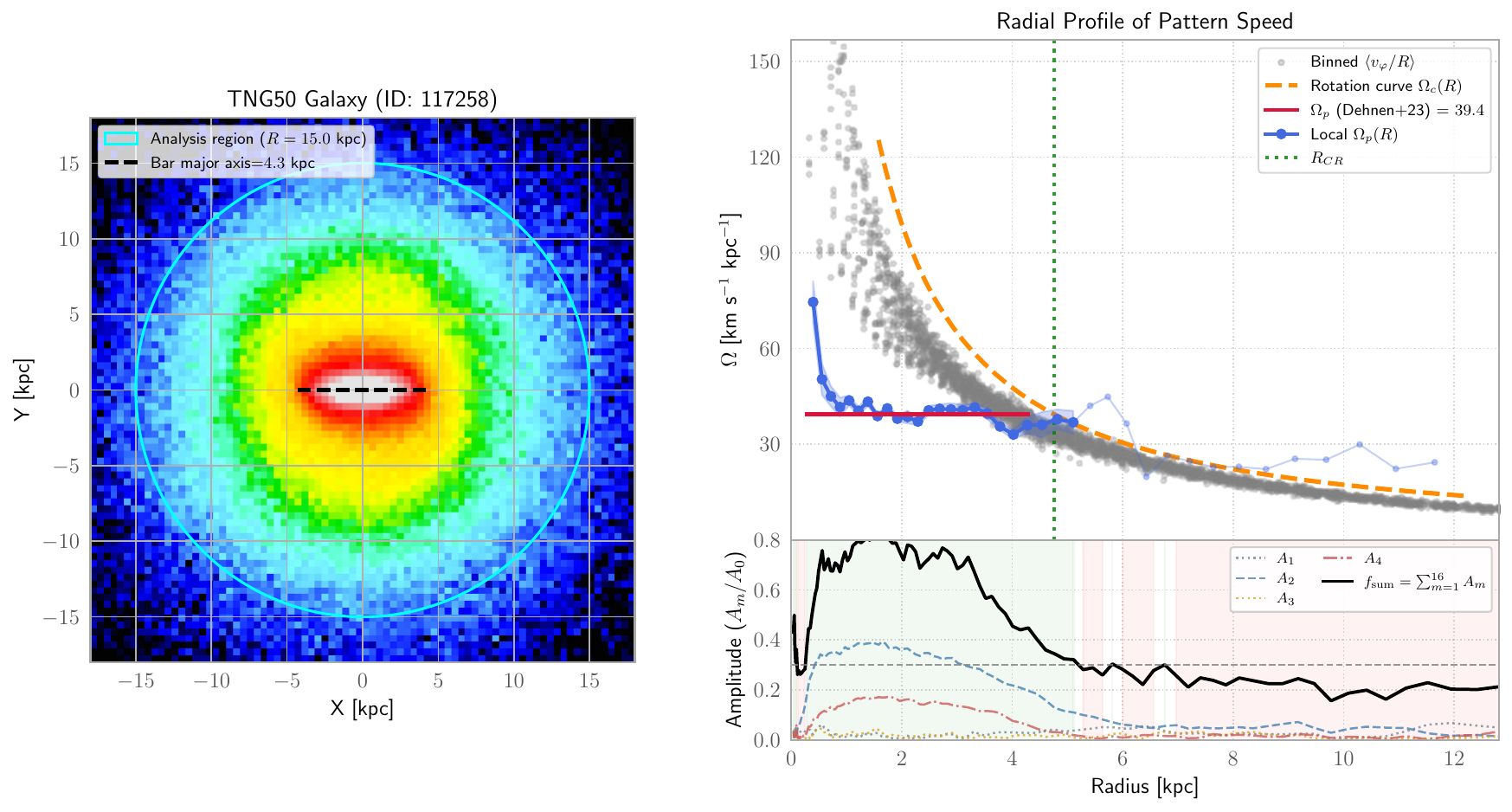}{0.7\textwidth}{(c) Ultrafast Candidate (ID 117258)}
\caption{Local pattern speed analysis for three distinct TNG50 barred galaxies. Each panel corresponds to a specific galaxy: (a) a barred galaxy with grand-design spiral arms, (b) a classic bar with no significant outer structures, and (c) a failed candidate for an ultrafast bar. The layout follows Figure~\ref{fig:radial_validation}, showing the radial profile of $\Omega_p(R)$ (blue solid line) and circular frequency $\Omega_c(R)$ (orange dashed line).}
\label{fig:three_bars}
\end{figure*}

In Figure~\ref{fig:three_bars}, we present the local pattern speed analysis for three distinct examples of barred galaxies from TNG50 at $z=0$, illustrating the diversity of kinematic structures captured by our method. These examples allow us to explore the relationship between the bar, spiral arms, and the fundamental dynamical limits of pattern rotation.

Figure~\ref{fig:three_bars}(a) (ID 560082) depicts a ``textbook" barred grand-design spiral galaxy. The local pattern speed profile reveals two distinct kinematic regimes. In the inner region ($R \lesssim 2$ kpc), $\Omega_{\rm p}(R)$ forms a flat plateau corresponding to the rigid rotation of the bar. Crucially, the method also successfully recovers a coherent signal in the outer disk ($R > 4$ kpc), coincident with the prominent spiral arms visible in the surface density map. The spiral structure exhibits a pattern speed that is distinct from the bar. If a spiral arm is a steady-state density wave, its pattern speed should manifest as a radial plateau. Conversely, a pattern speed that varies with radius suggests a transient structure that is winding up or dissolving. In this specific case, the outer structure shows a relatively flat profile, hinting at stability, though we explore the distinction between steady and transient spirals further in subsequent sections.

Figure~\ref{fig:three_bars}(b) (ID 429471) represents a classic barred galaxy lacking significant spiral features. The pattern speed profile is dominated by the central plateau, which aligns perfectly with the bar region. Beyond the corotation radius ($R_{\rm CR} \approx 10.5$ kpc), the non-axisymmetry strength $f_{\rm sum}$ drops significantly, and the measured pattern speed becomes consistent with the differential rotation of the disk or noise, as expected for an isolated bar.

Finally, we address the existence of ``ultrafast" bars in TNG50 using Figure~\ref{fig:three_bars}(c) (ID 117258). The rotation rate of a bar is often parameterized by the dimensionless distance ratio $\mathcal{R} \equiv R_{\rm CR}/R_{\rm bar}$, where $R_{\rm CR}$ is the corotation radius and $R_{\rm bar}$ is the bar semi-major axis. Theoretical dynamical arguments \citep{1980A&A....81..198C, 1980A&A....88..184A} demonstrate that a self-consistent weak bar cannot extend beyond its corotation radius, implying $\mathcal{R} > 1$. Furthermore, studies of dust lane shapes in barred galaxies predict $\mathcal{R} \approx 1.2 \pm 0.2$ \citep{1992MNRAS.259..345A}. The classification of bars into dynamical regimes based on the parameter $\mathcal{R} \equiv R_{\rm CR} / R_{\rm bar}$ has been a subject of evolving consensus.
Conventionally, bars are divided into ``fast'' ($1.0 < \mathcal{R} < 1.4$) and ``slow'' ($\mathcal{R} > 1.4$) categories.
Earlier major studies \citep[e.g.,][]{2008MNRAS.388.1803R, 2015A&A...576A.102A} found that the fast regime dominates the population, with slow bars being a minority.
However, this view has been challenged by recent analyses utilizing MaNGA data \citep{2019MNRAS.482.1733G, 2022MNRAS.517.5660G, 2023MNRAS.521.1775G}, which reported a higher fraction of slow bars and even identified a $\sim 10\%$ population of ``ultrafast'' candidates ($\mathcal{R} < 1.0$), although the physical reality of the latter remains contested \citep[see][]{2021A&A...649A..30C}.

However, recent analyses of TNG50 simulations suggest a tension with observations. \citet{2022ApJ...940...61F} noted that while the kinematic properties are physically consistent, TNG50 bars appear statistically shorter than observed counterparts for similar pattern speeds, thereby populating the ``slow" regime more frequently (higher $\mathcal{R}$ than observation). A similar discrepancy between simulation and observation is evident in Figure~8 of \citet{2024A&A...691A.122H}.

Our local pattern speed framework offers a direct visual diagnostic for the ``ultrafast" condition ($\mathcal{R} < 1$). If a bar were rotating faster than the local circular velocity at its tip, the plateau of the local pattern speed $\Omega_{\rm p}(R)$ would cross above the angular circular frequency curve $\Omega_c(R)$ (orange dashed line) within the bar region ($\Omega_p > \Omega_c$). In almost all analyzed galaxies in our sample—including the typical cases shown in Figures~\ref{fig:radial_validation}, \ref{fig:three_bars}(a), and \ref{fig:three_bars}(b)—the bar pattern speed remains strictly below the $\Omega_c(R)$ curve, satisfying $\mathcal{R} > 1$.

Figure~\ref{fig:three_bars}(c) represents one of the most extreme cases we identified, serving as a potential candidate for an ultrafast bar. The bar extends very close to its corotation radius. Yet, even in this limiting case, the local pattern speed plateau does not exceed the circular frequency curve within the bar extent; it merely approaches it. Consequently, despite the statistical tendency for TNG50 bars to be ``slow", we find no evidence for the breakdown of physical self-consistency (i.e., no ultrafast bars with $\mathcal{R} < 1$) within the simulation. This confirms that while TNG50 bars may be structurally shorter than observed ones, they adhere to the fundamental dynamical constraints expected for self-consistent stellar systems.

\subsection{Spiral Feature Diversity: From Density Waves to Material Arms}
\label{sec:spiral_diversity}

\begin{figure*}[ht!]
\centering
\fig{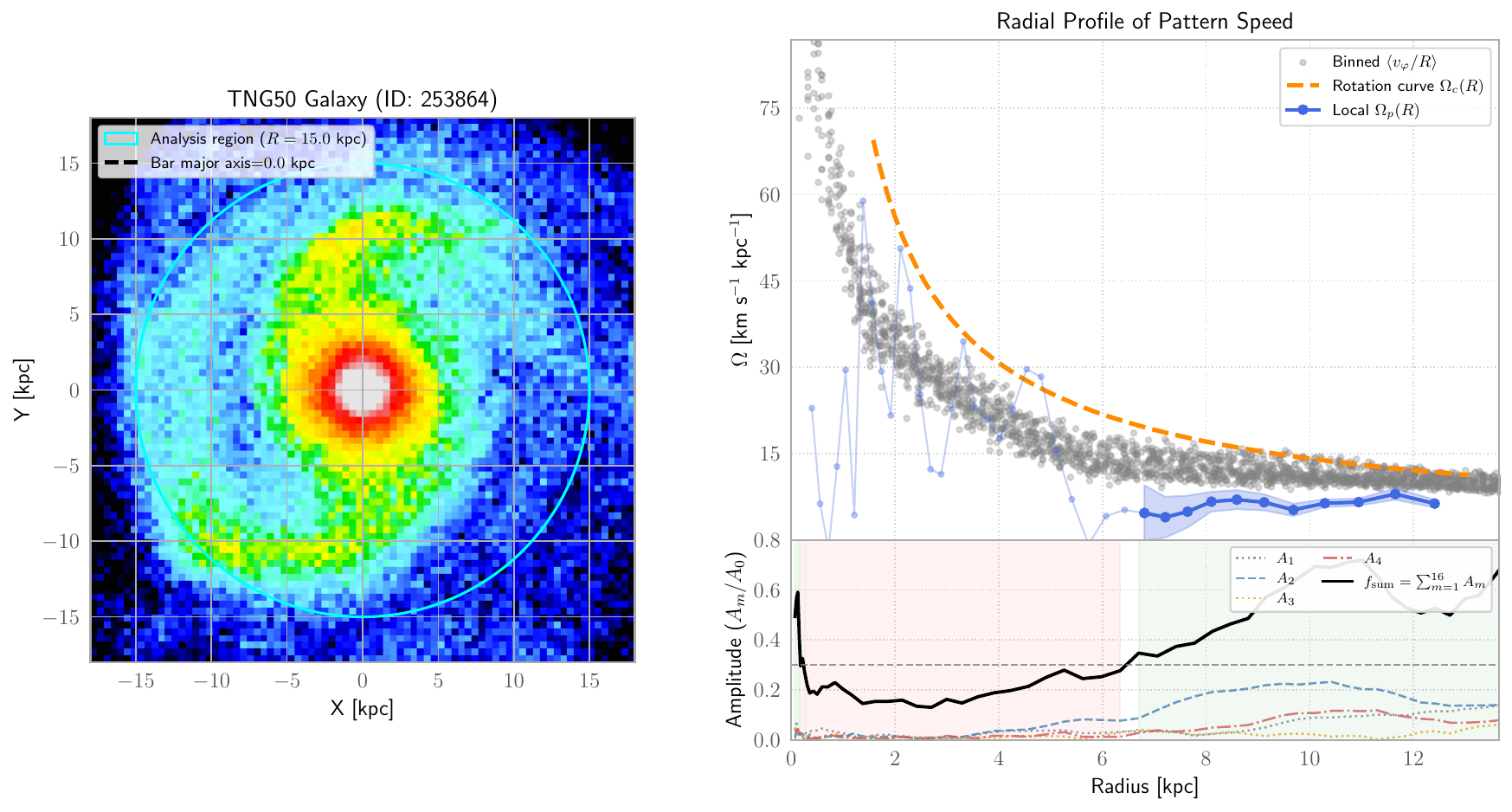}{0.7\textwidth}{(a) Density Wave Analog (ID 253864)}

\fig{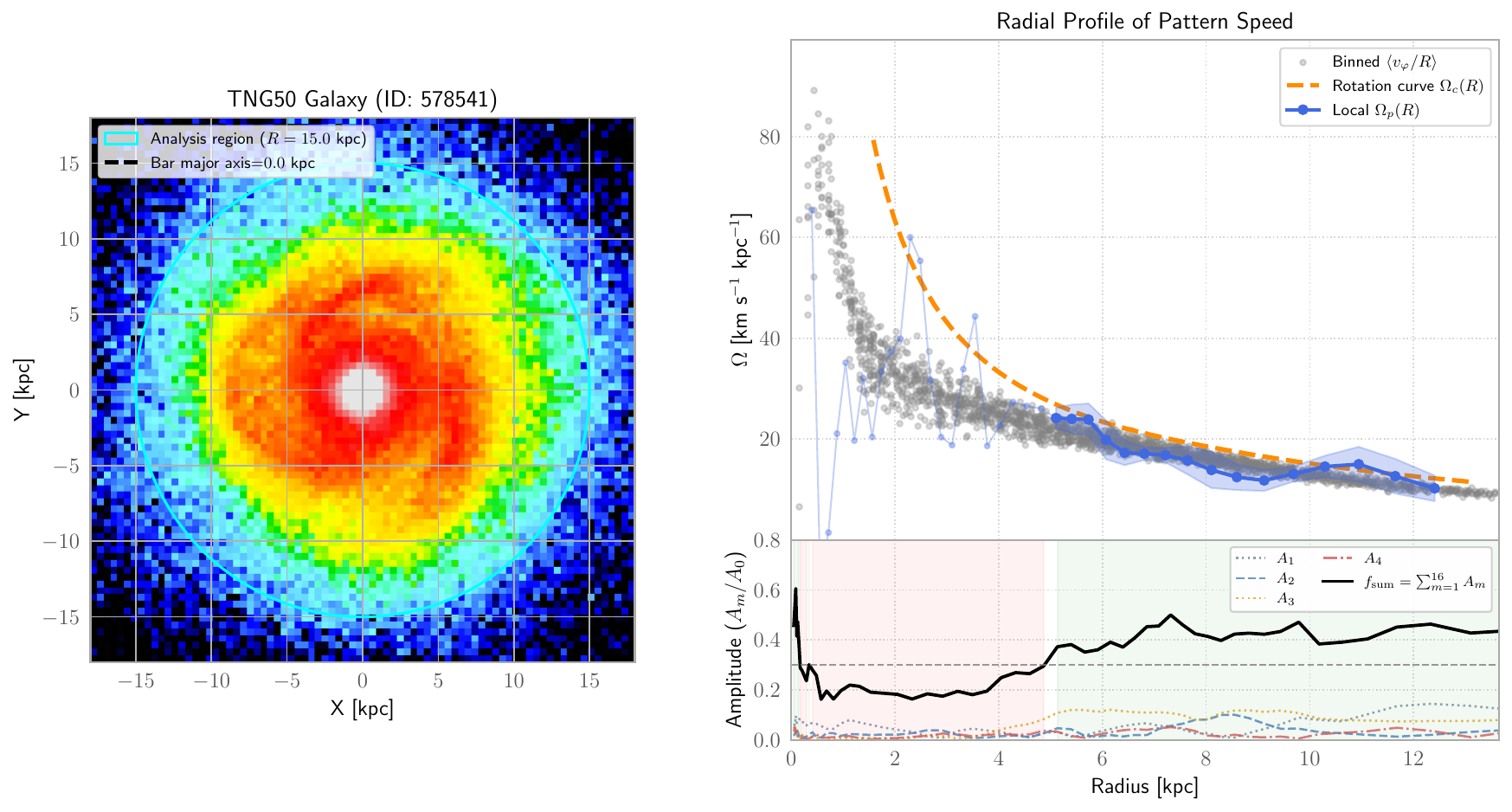}{0.7\textwidth}{(b) Material Arm / Co-rotating Spiral (ID 578541)}

\fig{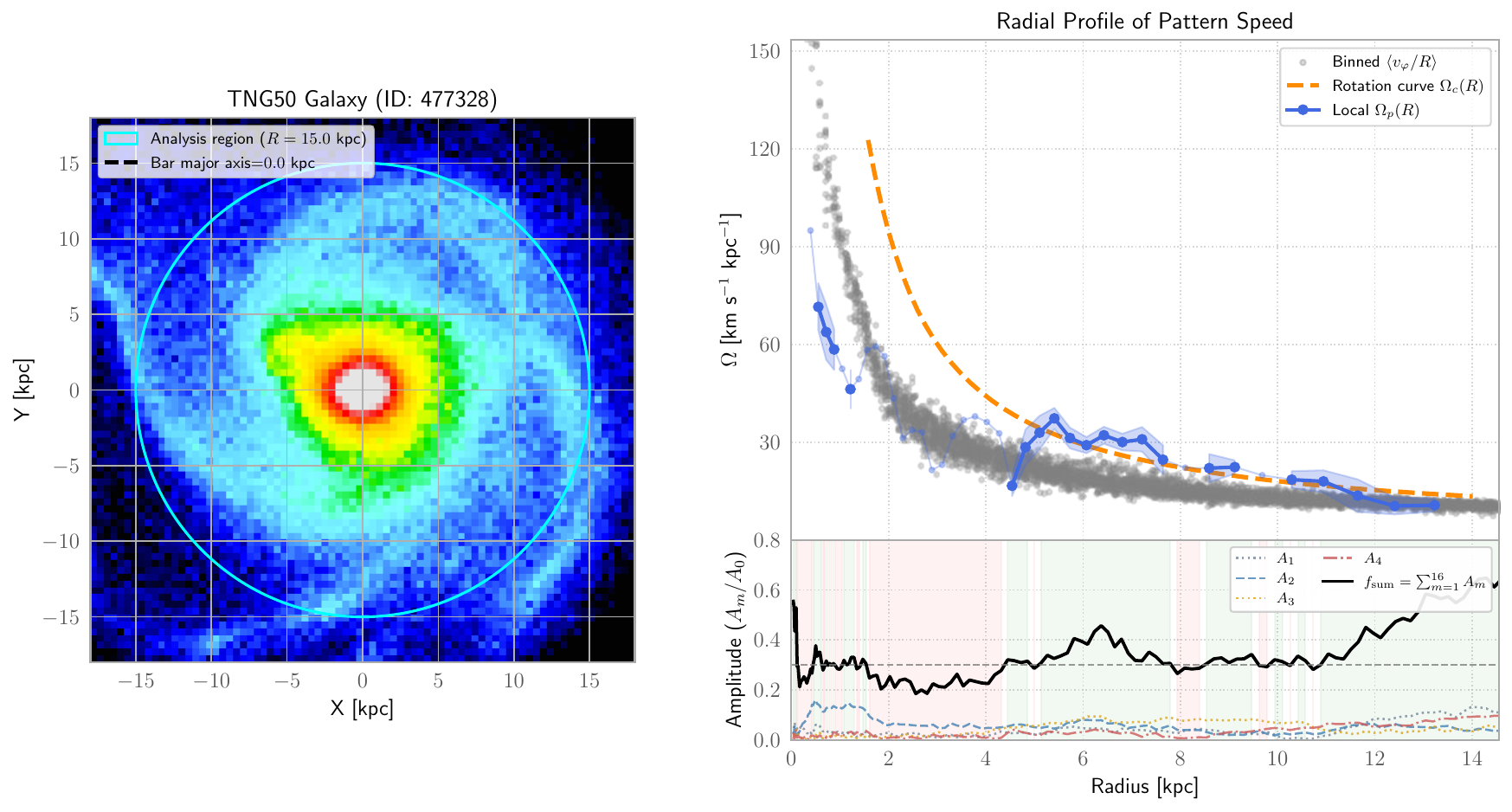}{0.7\textwidth}{(c) Fast-pattern Spiral (ID 477328)}
\caption{Local pattern speed analysis for three non-barred spiral galaxies in TNG50, illustrating three distinct dynamical regimes. The layout is identical to Figure~\ref{fig:radial_validation}. The grey points represent the binned tangential angular velocity of stellar particles, $\Omega_\varphi(R) = \langle v_{\varphi}/R \rangle$. (a) A two-armed spiral where the pattern speed (blue) forms a distinct plateau significantly slower than the stellar material speed. (b) A multi-armed spiral where the pattern speed tracks the differential rotation of the stars. (c) A complex spiral where the pattern speed locally exceeds the material speed.}
\label{fig:three_spirals}
\end{figure*}

While bars in TNG50 generally behave as coherent, rigid rotators obeying the expected dynamical limits (i.e., $\Omega_p < \Omega_c$), spiral arms exhibit a much richer diversity of kinematic behaviors. To explore this, we apply our local pattern speed framework to three non-barred spiral galaxies, identifying the dominant Fourier modes and comparing the derived pattern speeds $\Omega_p(R)$ with the local tangential velocity of stars $\langle v_{\varphi}/R \rangle$ (hereafter $\Omega_\varphi$) and the circular frequency $\Omega_c$. The results are compiled in Figure~\ref{fig:three_spirals}.

Figure~\ref{fig:three_spirals}(a) (ID 253864) highlights a galaxy with a prominent two-armed spiral structure (sometimes it would be misidentified visually as a barred galaxy). The Fourier analysis confirms the dominance of the $m=2$ mode (blue dashed line in the bottom panel) beyond $R=6$\,kpc. Crucially, in the radial range where the spiral is most prominent ($7 < R < 13$\,kpc), the measured local pattern speed $\Omega_p(R)$ manifests as a remarkably flat plateau ($\approx 5-8$ km s$^{-1}$ kpc$^{-1}$). This pattern speed is significantly lower than both the circular frequency curve $\Omega_c$ (orange dashed line) and, more importantly, the actual tangential angular velocity of the stars $\Omega_\varphi$ (grey dots). The clear decoupling between the pattern speed and the material speed ($\Omega_p \ll \Omega_\varphi$) is the hallmark of a classic quasi-stationary density wave \citep{1964ApJ...140..646L}. In this regime, stars flow through the spiral potential, maintaining the spiral structure as a standing wave pattern rather than a collection of fixed stars.

In contrast, Figure~\ref{fig:three_spirals}(b) (ID 578541) presents a flocculent, multi-armed morphology. The Fourier decomposition reveals a dominance of the $m=3$ mode in the outer disk ($R > 5$\,kpc). The kinematic profile here is fundamentally different from the previous case. The pattern speed profile does not form a flat plateau; instead, it declines with radius, closely tracking the differential rotation of the stellar material ($\Omega_p(R) \approx \Omega_\varphi(R)$). When the pattern rotation frequency matches the angular velocity of the stars, the structure behaves more akin to a ``material arm" or ``streaming motion" rather than a wave propagating through the medium. Such kinematic alignment is often indicative of transient, recurrent spirals formed via swing amplification or shearing instabilities, which tend to wind up and dissolve over dynamical timescales \citep{1981seng.proc..111T, 1984ApJ...282...61S}. The local pattern speed method effectively captures this behavior, distinguishing it from the rigid rotation seen in bars or grand-design density waves.

Figure~\ref{fig:three_spirals}(c) (ID 477328) introduces a third, more exotic category. This galaxy displays a complex four-armed structure, though the Fourier spectrum indicates a dominant $m=3$ component. Unlike the bar examples in Section~\ref{sec:typical_bars}, where $\Omega_p$ is strictly bounded by the stellar circulation, the spiral pattern speed in this galaxy exhibits anomalous high-velocity features. In several radial bins, $\Omega_p(R)$ exceeds not only the local tangential velocity $\Omega_\varphi$ but also approaches or exceeds the circular frequency $\Omega_c$. This indicates that the spiral pattern is propagating faster than the local stellar flow.

These three examples underscore a critical distinction between bars and spirals in a cosmological context. While bars are distinct, long-lived structural components that rotate slower than the co-spatial stars ($\Omega_p < \Omega_\varphi$), spiral arms in TNG50 display a full spectrum of dynamical states: from slow density waves, to co-rotating material features, to fast-moving patterns. The mechanisms driving these ``super-rotation" features in spirals (Figure~\ref{fig:three_spirals}c) likely differ from the secular evolution driving bar slowdowns. While a full analysis of the formation mechanisms for these diverse spiral types is beyond the scope of this paper, our results demonstrate that radially resolved pattern speed measurements are essential for classifying the physical nature of spiral arms in large-scale simulations.

\subsection{Complex Dynamical States: Interactions, Coupled Structures, and Hidden Bars}
\label{sec:complex_states}

\begin{figure*}[ht!]
\centering
\fig{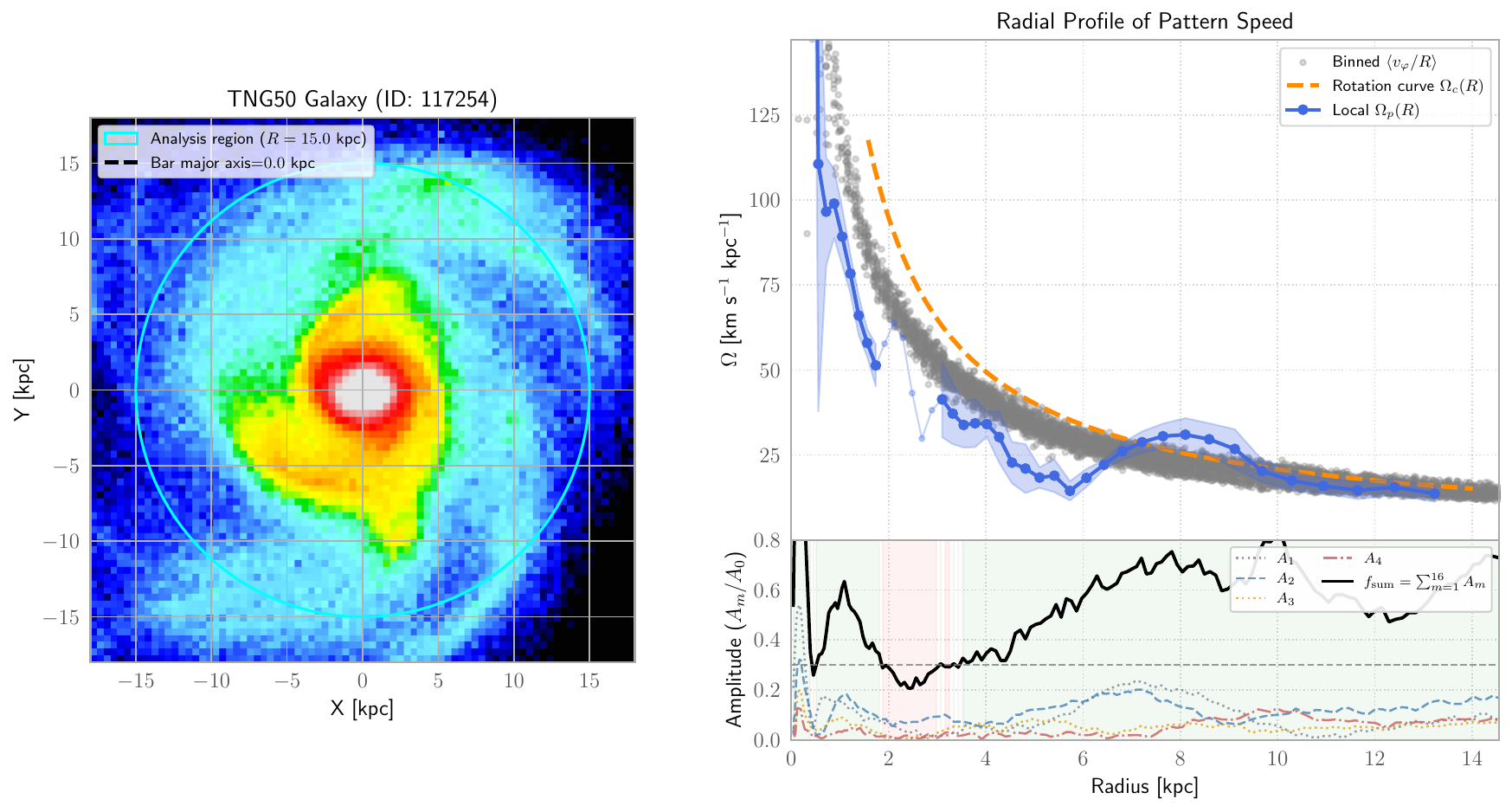}{0.7\textwidth}{(a) Tidally Disturbed Galaxy (ID 117254)}

\fig{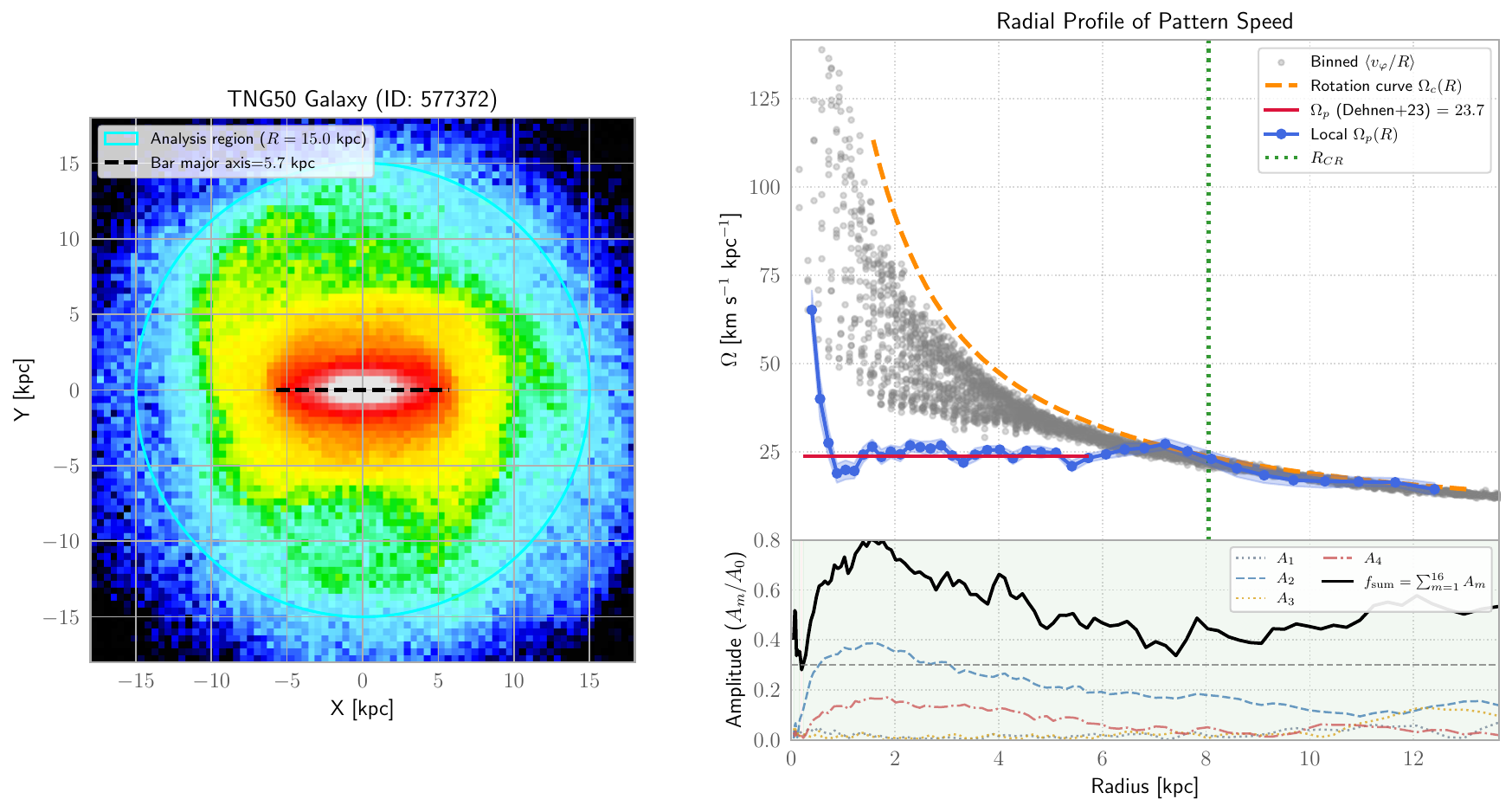}{0.7\textwidth}{(b) Bar Coupled with Material Arms (ID 577372)}

\fig{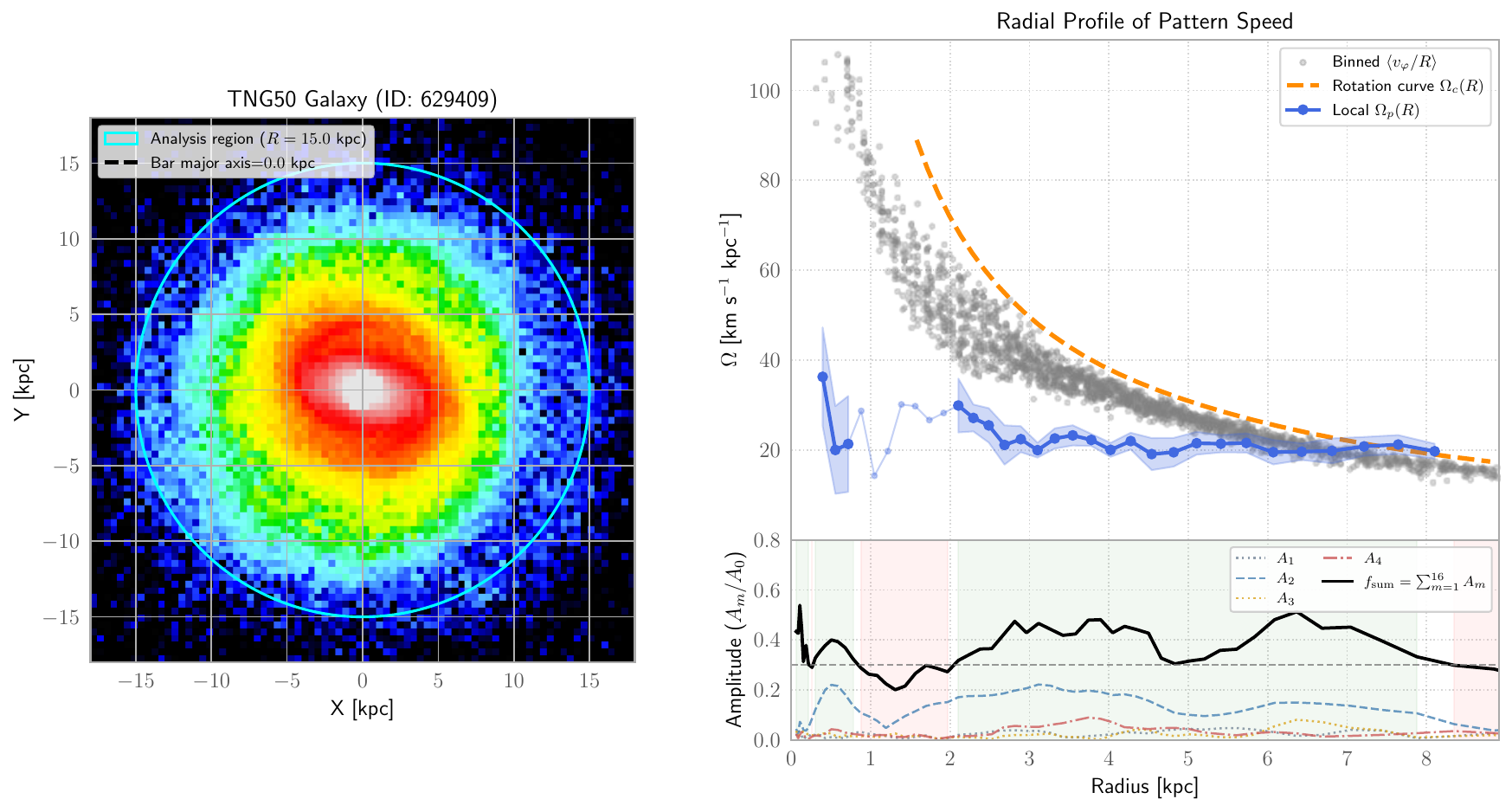}{0.7\textwidth}{(c) The Kinematic ``Hidden'' Bar/Weak Oval Distortion (ID 629409)}
\caption{Local pattern speed analysis illustrating the method's capability to diagnose complex dynamical states. The layout mirrors Figure~\ref{fig:three_bars}. 
(a) A galaxy undergoing a flyby interaction or merger remnant phase. Note the strong $m=1$ mode (bottom panel, dotted black) and the ``super-rotating" feature at $R\sim8$\,kpc where $\Omega_p > \Omega_c$. 
(b) A strong bar connected to a two-armed spiral. The bar rotates as a rigid body ($R<6$\,kpc), while the connected outer arms shear with the differential rotation of the disk ($R>8$\,kpc). 
(c) A galaxy with no visually identifiable bar and weak non-axisymmetry ($A_2 < 0.2$), yet revealing a distinct, flat pattern speed plateau separate from the stellar streaming motion, indicative of a weak rotating oval distortion.}
\label{fig:exotic_cases}
\end{figure*}

The cosmological environment of TNG50 naturally gives rise to galaxies that defy simple classification into ``stable bars" or ``steady density waves." Galaxies undergo flyby interactions, possess multi-component structures with distinct rotation rates, or harbor weak non-axisymmetric potentials that are visually obscure. In this subsection, we demonstrate the diagnostic power of the local pattern speed framework in these complex regimes. We present three distinct cases in Figure~\ref{fig:exotic_cases}, covering a tidal interaction, a bar-spiral system, and a ``hidden" bar.

Figure~\ref{fig:exotic_cases}(a) (ID 117254) illustrates a system far from equilibrium, characterized by severe morphological lopsidedness. The Fourier decomposition (bottom panel) highlights a dominant $m=1$ mode (black dotted line), typically associated with tidal interactions or lopsided instability. Unlike the relaxed disks discussed previously, the ``pattern speed" here largely tracks the tangential velocity of the stars, suggesting that the non-axisymmetry is material in nature rather than a wave. However, a striking anomaly appears between $7 < R < 9$\,kpc: the measured pattern speed locally exceeds the circular frequency curve ($\Omega_p > \Omega_c$). In a steady-state isolated galaxy, this would violate theoretical limits. However, in the context of a flyby, this feature likely traces a coherent stream of stripped stars or a tidal tail with high specific angular momentum that is phase-wrapping around the main galaxy. Our method successfully flags this region as dynamically distinct, revealing footprints of interaction that might be missed by global averaging techniques.

Figure~\ref{fig:exotic_cases}(b) (ID 577372) presents a scenario where a strong bar is visually connected to a grand-design two-armed spiral. A key question in galactic dynamics is whether such spirals propagate at the same speed as the bar (manifold-driven) or decouple. The radial profile provides an unambiguous answer. The inner region ($R<6$\,kpc) exhibits a classic, flat bar plateau ($\Omega_p \approx 23$\,km s$^{-1}$ kpc$^{-1}$). Crucially, at the spiral transition region ($R > 8$\,kpc), the pattern speed drops abruptly and adheres closely to the stellar material curve ($\Omega_p \approx \Omega_\varphi$), declining with radius. This confirms that while the spiral arms may be triggered by the bar, they behave kinematically as transient material arms winding up with differential rotation, distinct from the rigid rotation of the bar itself. This decoupling of bar and spiral pattern speeds within a single galaxy mirrors the statistical results found in observations \citep[e.g.,][]{1988MNRAS.231P..25S} but is here resolved spatially for a single object.

Finally, in Figure~\ref{fig:exotic_cases}(c) (ID 629409), we examine a galaxy that would typically be classified as non-barred. Visually, the galaxy appears as an amorphous disk-bulge system, and the normalized Fourier amplitude of the bi-symmetry mode remains below the conventional threshold for bar detection ($\Sigma_2/\Sigma_0 < 0.2$) across the entire disk. Nevertheless, our kinematic analysis uncovers a ``hidden" structural component. The local pattern speed profile reveals a coherent, flat plateau at $\Omega_p \approx 20$\,km s$^{-1}$ kpc$^{-1}$ extending out to $R \approx 5$\,kpc. This rotation is significantly slower than, and decoupled from, the local tangential velocity of the stars (grey points). This signature implies the presence of a weak, triaxial oval distortion or a ``ghost" bar—a structure dynamically coherent enough to possess a pattern speed but too faint to trigger morphological selection criteria. This highlights a unique strength of the local pattern speed framework: the ability to identify dynamical structures based on their phase coherence rather than their density contrast alone.

\section{Discussion and Conclusions} 
\label{sec:discussion_conclusions}

In this paper, we presented a general integral form of the continuity equation as a framework for measuring pattern speeds in galaxies. By transforming the problem from global integration to local flux conservation, we provide a flexible, spatially resolved tool for probing the internal kinematics of non-axisymmetric structures. 

Applying the local pattern speed method to a statistical sample of barred galaxies in the TNG50 simulation, we found that bars are characterized by a distinct, constant pattern speed plateau. Notably, we find no dynamical evidence for ``ultrafast" bars ($\Omega_p > \Omega_c$), where the pattern speed exceeds the local circular frequency. Furthermore, unlike the rigid rotation characteristic of bars, spiral arms in TNG50 display a rich diversity of kinematic behaviors. Our method successfully distinguishes between: (i) classical density waves traveling significantly slower than the stellar fluid ($\Omega_p < \Omega_\varphi$); (ii) material arms that co-rotate with the differential shearing of the disk ($\Omega_p \approx \Omega_\varphi$); and (iii) fast-propagating patterns. Our framework also allows for the detection of dynamical features that are often obscured, such as ``hidden" bars (weak oval distortions) in galaxies with no obvious photometric bar.

Beyond presenting new structural measurements, the concept of a ``local pattern speed" provides a highly intuitive perspective on the traditional Tremaine-Weinberg (TW) method and its variations. Reinterpreting historical methods through this lens introduces several distinct advantages:

\begin{enumerate}
    \item \textbf{Clear physical intuition:} As depicted in Figures~\ref{fig:local-geo} and \ref{fig:unified-geo}, our approach complements derivations based on differential operators by offering a visually intuitive, geometric alternative for understanding the underlying mechanics of pattern propagation. 

    \item \textbf{Methodological unification:} Our framework facilitates the comparison and categorization of various established methods. As discussed in Section~\ref{sec:tw_special_cases}, the celebrated formula of \citet{1984ApJ...282L...5T}, the proper-motion-based method for the Milky Way bar \citep{2019MNRAS.488.4552S}, and the line-of-sight velocity approach for local structures \citep[e.g.,][]{2002MNRAS.334..355D} can all be mathematically contextualized simply by selecting appropriate integration loops. This generalization clarifies the geometric conditions and analytical assumptions under which each method is most effective, as explicitly highlighted in our analysis of the traditional radial TW slit formulation (Section~\ref{sec:radial_tw} and Figure~\ref{fig:radial_tw_approx}).

    \item \textbf{Enhanced diagnostic power:} Our analysis clarifies that measurements over regions containing multiple overlapping patterns yield an average speed weighted by the strength of the non-axisymmetric signature. Consequently, the radially resolved ``local pattern speed" profile serves as a powerful diagnostic: constant plateaus indicate coherent, stable patterns (like bars or grand-design spirals), while radially varying profiles trace transient material features or differential rotation.
\end{enumerate}

A concrete implementation of this formalism is presented in the companion paper \citet{2026arXiv260321279D}, where we apply the local pattern speed concept to analyze the excess of bar-like structures in TNG50, demonstrating its efficacy in capturing galactic bar features.

Ultimately, treating pattern speed measurement as a localized, spatial filter provides an unprecedented opportunity to dissect complex galactic dynamics. Because real galaxies frequently host multiple overlapping patterns, the act of carefully selecting the integration loop allows one to isolate specific structures—for instance, using a compact loop to isolate a central bar, while employing an outer annular loop to target the distinct kinematics of spiral arms. Building upon the versatility of this framework, in future research we will investigate the specific formation mechanisms and evolutionary drivers of the diverse spiral types identified in this study.

\begin{acknowledgments}
We thank Youjun Lu and Shude Mao for their expert guidance and valuable discussions that greatly enhanced this work. We are also indebted to E. Athanassoula for insightful discussions during the early stages of the project, which significantly helped refine the conceptual design of the study. We thank the anonymous referee for their constructive comments. This work is supported by by the National Key Research and Development Program of China (No. 2023YFA1607904), the National Astronomical Observatories of the Chinese Academy of Sciences (No. E4ZR0510), the Beijing Municipal Natural Science Foundation (No. 1242032), and the Youth Innovation Promotion Association of the Chinese Academy of Sciences (No. 2022056). The IllustrisTNG simulations were undertaken with compute time awarded by the Gauss Centre for Supercomputing (GCS) under GCS Large-Scale Projects GCS-ILLU and GCS-DWAR on the GCS share of the supercomputer Hazel Hen at the High Performance Computing Center Stuttgart (HLRS), as well as on the machines of the Max Planck Computing and Data Facility (MPCDF) in Garching, Germany. We are grateful for the use of the TNG project's online JupyterLab service and high-performance computing resources, which greatly facilitated the data analysis presented in this work.
\end{acknowledgments}



\appendix

\section{Derivation for Arbitrary Geometries}
\label{sec:general_derivation}

Let us define the infinitesimal line element along the closed loop as $d\mathbf{l}$, with the positive direction defined counter-clockwise. The outward-pointing normal vector unit is $\hat{\mathbf{n}}$. By observing the structure of Equation~\eqref{eq:local_pattern_speed}, we can identify the numerator as the total circulation of mass flux and the denominator as a moment of the density distribution. Generalizing the polar velocity components to vector products, we propose the following general integral form:
\begin{equation}
    \Omega_{\rm p} = \frac{\oint_{\partial S} \Sigma \mathbf{v} \cdot (\hat{\mathbf{z}} \times d\mathbf{l})}{\oint_{\partial S} \Sigma \mathbf{r} \cdot d\mathbf{l}}.
    \label{eq:heuristic_general}
\end{equation}
Recognizing that the term $\mathbf{v} \cdot (\hat{\mathbf{z}} \times d\mathbf{l})$ is equivalent to $-v_n dl$ (where $v_n = \mathbf{v} \cdot \hat{\mathbf{n}}$ is the velocity component normal to the boundary), we can rewrite this in a more compact form:
\begin{equation}
    \Omega_{\rm p} = -\frac{\oint_{\partial S} \Sigma v_n \, dl}{\oint_{\partial S} \Sigma \mathbf{r} \cdot d\mathbf{l}}.
\end{equation}
This equation suggests that the pattern speed is simply the ratio of the net normal mass flux out of the region to the first moment of the density along the boundary. 

We can rigorously validate the heuristic generalizations in Equations~\eqref{eq:heuristic_general} and \eqref{eq:general_loop_integral} by starting from the continuity equation. This derivation ensures that our geometric intuition holds firm when subjected to the strict requirements of hydrodynamics.

In a reference frame rotating with the pattern at angular velocity $\Omega_{\rm p}$, a steady-state wave requires that the surface density $\Sigma$ is stationary in time ($\partial \Sigma / \partial t = 0$). The continuity equation in this rotating frame is given by \citet[p. 475]{2008gady.book.....B}:
\begin{equation}
    \nabla \cdot \left[ \Sigma (\mathbf{v} - \Omega_{\rm p} \hat{\mathbf{z}} \times \mathbf{r}) \right] = 0.
\end{equation}
Here, the term in the brackets represents the mass flux relative to the rotating pattern. Rearranging this equation to isolate the terms involving $\Omega_{\rm p}$, we obtain:
\begin{equation}
    \Omega_{\rm p} \left[ \nabla \cdot (\Sigma \hat{\mathbf{z}} \times \mathbf{r}) \right] = \nabla \cdot (\Sigma \mathbf{v}).
    \label{eq:continuity_rotating_expanded}
\end{equation}
Equation~\eqref{eq:continuity_rotating_expanded} acts as the local differential form of the pattern speed definition. To retrieve the integral form, we integrate both sides over an arbitrary surface area $S$ bounded by the closed loop $\partial S$ and apply the divergence theorem (Gauss's Theorem).

The integration proceeds as follows:
\begin{align}
    \Omega_{\rm p} &= \frac{\int_S \nabla \cdot (\Sigma \mathbf{v}) \, dS}{\int_S \nabla \cdot (\Sigma \hat{\mathbf{z}} \times \mathbf{r}) \, dS} \nonumber \\
    &= \frac{\oint_{\partial S} \Sigma \mathbf{v} \cdot \hat{\mathbf{n}} \, dl}{\oint_{\partial S} (\Sigma \hat{\mathbf{z}} \times \mathbf{r}) \cdot \hat{\mathbf{n}} \, dl}.
    \label{eq:line_integral_balance}
\end{align}
We now simplify the denominator. Note that on the boundary curve, $\hat{\mathbf{n}} \, dl = - \hat{\mathbf{z}} \times d\mathbf{l}$ (assuming counter-clockwise integration). Utilizing the scalar triple product identity $(\mathbf{A} \times \mathbf{B}) \cdot \mathbf{C} = (\mathbf{B} \times \mathbf{C}) \cdot \mathbf{A}$, the denominator becomes:
\begin{equation}
    (\Sigma \hat{\mathbf{z}} \times \mathbf{r}) \cdot \hat{\mathbf{n}} \, dl 
    = \Sigma (\hat{\mathbf{z}} \times \mathbf{r}) \cdot (-\hat{\mathbf{z}} \times d\mathbf{l})
    = -\Sigma \mathbf{r} \cdot d\mathbf{l}.
\end{equation}
Substituting this back into Equation~\eqref{eq:line_integral_balance}, and recalling that $\mathbf{v} \cdot \hat{\mathbf{n}} = v_n$, we arrive at the final general expression:
\begin{equation}
    \Omega_{\rm p} = \frac{\oint_{\partial S} \Sigma \mathbf{v} \cdot (-\hat{\mathbf{z}} \times d\mathbf{l})}{\oint_{\partial S} (\hat{\mathbf{z}} \times \Sigma \mathbf{r}) \cdot (-\hat{\mathbf{z}} \times d\mathbf{l})} 
    = -\frac{\oint_{\partial S} \Sigma v_n \, dl}{\oint_{\partial S} \Sigma \mathbf{r} \cdot d\mathbf{l}}.
    \label{eq:local_omega_p_general}
\end{equation}
This confirms that Equation~\eqref{eq:local_omega_p_general} is identical to our heuristic Equation~\eqref{eq:general_loop_integral}. The pattern speed is strictly determined by the line integrals of the tracer's normal velocity and position along the region's boundary.

\section{Derivation for Multiple Pattern Speeds}
\label{sec:multiple_derivation}

Let us return to the simple geometry of an annular sector, as illustrated in Figure~\ref{fig:multiple-patterns}. We consider an integration region spanning radii from $r_0$ to $r_N$ and azimuths from $\varphi_1$ to $\varphi_2$. We conceptually divide this region into $N$ narrow, contiguous sub-annuli, where the $n$-th annulus extends from $r_n$ to $r_{n+1}$. We assume that within each sufficiently narrow sub-annulus, the pattern speed can be treated as approximately constant, denoted by $\Omega_{\rm p}(r_n)$.

For each individual sub-annulus, the condition of mass conservation must hold. The balance between the pattern-induced mass change and the net physical flux across its boundaries can be written based on Equation~(\ref{eq:local_pattern_speed}):
\begin{equation}
    \Omega_{\rm p}(r_n) \int_{r_n}^{r_{n+1}} \left[ \Sigma(r, \varphi_1) - \Sigma(r, \varphi_2) \right] r \, dr = \mathcal{F}_n,
    \label{eq:flux_balance_subannulus}
\end{equation}
where $\mathcal{F}_n$ represents the total net flux of the tracer out of the $n$-th sub-annulus:
\begin{equation}
    \mathcal{F}_n = \left[ \int_{r_n}^{r_{n+1}} \Sigma v_\varphi \, dr \right]_{\varphi_1}^{\varphi_2} + \left[ \int_{\varphi_1}^{\varphi_2} \Sigma v_r r \, d\varphi \right]_{r_n}^{r_{n+1}}.
\end{equation}
Here, the notation $\left. f \right|_{a}^{b}$ again denotes $f(b) - f(a)$.

Now, let us sum Equation~(\ref{eq:flux_balance_subannulus}) over all $N$ sub-annuli, from $n=0$ to $N-1$. On the right-hand side, the sum of the fluxes, $\sum_{n=0}^{N-1} \mathcal{F}_n$, telescopes. The radial flux across a shared boundary $r_n$ from adjacent annuli cancels out, leaving only the flux across the outermost boundaries of the entire region (at $r_0$, $r_N$, $\varphi_1$, and $\varphi_2$). The sum therefore yields the total flux, $\mathcal{F}_{\rm total}$, out of the full annular sector:
\begin{align}
    \sum_{n=0}^{N-1} \mathcal{F}_n &= \left[ \int_{r_0}^{r_N} \Sigma v_\varphi \, dr \right]_{\varphi_1}^{\varphi_2} + \left[ \int_{\varphi_1}^{\varphi_2} \Sigma v_r r \, d\varphi \right]_{r_0}^{r_N} \equiv \mathcal{F}_{\rm total}.
    \label{eq:summed_full_annular}
\end{align}
Equating this to the sum of the left-hand sides of Equation~(\ref{eq:flux_balance_subannulus}), we get:
\begin{equation}
    \sum_{n=0}^{N-1} \left( \Omega_{\rm p}(r_n) \int_{r_n}^{r_{n+1}} \left[ \Sigma(r, \varphi_1) - \Sigma(r, \varphi_2) \right] r \, dr \right) = \mathcal{F}_{\rm total}.
    \label{eq:summed_balance}
\end{equation}

Now, compare this result to what is measured if we naively apply our original formula (Eq.~\ref{eq:local_pattern_speed}), which assumes a single pattern speed, to the entire region. The measured value, $\langle \Omega_{\rm p} \rangle$, would be defined by:
\begin{equation}
    \langle \Omega_{\rm p} \rangle \int_{r_0}^{r_N} \left[ \Sigma(r, \varphi_1) - \Sigma(r, \varphi_2) \right] r \, dr = \mathcal{F}_{\rm total}.
    \label{eq:naive_measurement}
\end{equation}
Since the right-hand sides of Equations~(\ref{eq:summed_balance}) and (\ref{eq:naive_measurement}) are identical, we can equate their left-hand sides and solve for the measured pattern speed $\langle \Omega_{\rm p} \rangle$:
\begin{equation}
    \langle \Omega_{\rm p} \rangle = \frac{\sum_{n=0}^{N-1} \Omega_{\rm p}(r_n) \cdot w_n}{\sum_{n=0}^{N-1} w_n},
\end{equation}
where the weight $w_n$ for each sub-annulus is
\begin{equation}
    w_n = \int_{r_n}^{r_{n+1}} \left[ \Sigma(r, \varphi_1) - \Sigma(r, \varphi_2) \right] r \, dr.
\end{equation}

This result implies that the measured pattern speed is biased towards regions with strong non-axisymmetric features. We note that \citet{2002MNRAS.334..355D} anticipated this behavior. In their analysis of potential complications (their Section 3.3), they argued based on the linearity of the continuity equation that the measured value would be a weighted average, $\overline{\Omega} \propto \sum \mathcal{K}_i / \sum \mathcal{P}_i$, where $\mathcal{K}$ and $\mathcal{P}$ are kinematic and photometric integrals. Our general loop-integral derivation provides the rigorous geometric proof of this intuition, explicitly defining the weight $w_n$ (Eq.~\ref{eq:weight_definition}) in terms of the density contrast along the integration path.

This is a central result of our framework. It formally demonstrates that any measurement over a finite region with varying pattern speeds yields a weighted average. The nature of the weight, $w_n$, is physically illuminating. It is not simply the mass within the sub-annulus, but rather an integral of the azimuthal density difference, $\Sigma(r, \varphi_1) - \Sigma(r, \varphi_2)$, across the sector. This quantity is a first-order moment of the non-axisymmetric surface density; it measures the extent to which the tracer distribution deviates from axisymmetry within that radial bin.

This has profound practical implications. The measured $\langle \Omega_{\rm p} \rangle$ is biased towards the pattern speeds of regions where the non-axisymmetric signal of the tracer is strongest along the chosen integration path. A dynamically prominent but nearly axisymmetric disk component will have $w_n \approx 0$ and will contribute minimally to the measured average speed, even if it harbors a weak, fast pattern. Conversely, a less massive but strongly barred or spiral region will dominate the weighted average.

This result, derived here for a polar grid, generalizes directly to the arbitrary loop formulation of Equation~(\ref{eq:local_omega_p_general}). In the general case, the measured pattern speed is the average of the local $\Omega_{\rm p}(\mathbf{x})$ weighted by the ``pattern flux density", $\Sigma ((\hat{\mathbf{z}} \times \mathbf{r}) \cdot \hat{\mathbf{n}})$, integrated along the loop $\partial S$.

This understanding allows for a more nuanced interpretation of pattern speeds measured with any TW-type method. The seemingly single value of $\Omega_{\rm p}$ obtained from a global measurement is, in fact, a complex luminosity- and geometry-weighted average of all patterns present. This insight also opens a new investigative avenue: by carefully designing the integration loop $\partial S$, one can hope to preferentially weight certain regions of a galaxy to isolate the pattern speeds of different components, such as the bar versus the spiral arms. This concept forms the basis for the refined method for measuring radially varying pattern speeds that we develop in Section~\ref{sec:radial_tw}.

\section{Derivation for Theoretical Unification}
\label{sec:tw_special_cases}

\subsection{The Classic TW Method for External Galaxies}
\label{sec:classic_tw}

The classic Tremaine-Weinberg method is arguably the most fundamental and widely used model-independent technique for measuring bar pattern speeds in external galaxies \citep{1984ApJ...282L...5T}. It has been successfully applied to many galaxies using stellar kinematics \citep[e.g.,][]{2015A&A...576A.102A, 2019A&A...632A..51C, 2019MNRAS.482.1733G, 2020MNRAS.491.3655G, 2022MNRAS.517.5660G, 2020A&A...641A.111C, 2021A&A...649A..30C, 2021AJ....161..185W, 2023MNRAS.521.1775G}. We will now show that this seminal method emerges directly from our general integral formulation by choosing a specific, albeit idealized, integration path.

\begin{figure}[ht!]
    \centering
    \includegraphics[width=0.6\textwidth]{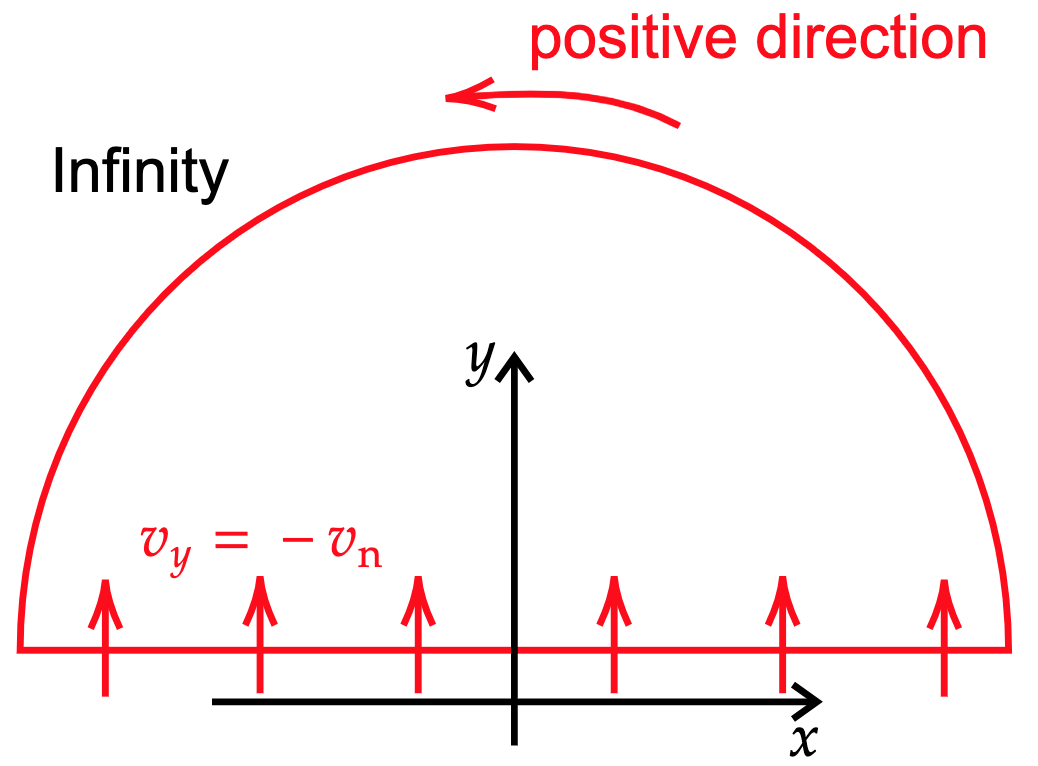} 
    \caption{Schematic of the closed integration loop used to derive the classic Tremaine-Weinberg (TW) method from our general integral formalism. The path $\partial S$ (red line) consists of a straight segment along the galaxy's major axis ($x$-axis) and a large semicircular arc. The arrows along the straight segment represent the component of the velocity perpendicular to the integration path, $v_y$, which corresponds to the negative projection of the velocity onto the outward normal vector (-$\mathbf{v} \cdot \hat{\mathbf{n}}$) in this geometry (and equals to the observed line-of-sight velocity, $v_{\text{LOS}}/\sin i$, where $i$ is the galaxy inclination). In the idealized limit where the arc's radius approaches infinity, the tracer density $\Sigma$ is assumed to vanish, causing the line integrals along the arc to become zero. This isolates the contributions from the straight segment, which is analogous to an observational long slit. This specific geometry formally demonstrates how our general expression (Eq.~\ref{eq:local_omega_p_general}) recovers the classic TW formula (Eq.~\ref{eq:classic_tw}). In practice, observations use a finite-length slit, meaning the method implicitly measures the \textit{local} pattern speed within the region enclosed by the slit.
    }
    \label{fig:classic_tw_loop}
\end{figure}

Let us consider a galactic disk observed at an inclination $i$, with its major axis aligned with the $x$-axis of a Cartesian coordinate system centered on the galaxy's nucleus. The classic TW method utilizes kinematic and photometric data along one or more slits parallel to the major axis. For our derivation, we choose an integration loop $\partial S$ as depicted in Figure~\ref{fig:classic_tw_loop}. The loop consists of a straight line segment along the $x$-axis from $x=-L$ to $x=+L$, and a large semicircular arc of radius $L$ in the $y>0$ half-plane that closes the path.

We apply our general formula, Equation~(\ref{eq:local_omega_p_general}), to this loop. Let's evaluate the numerator and denominator separately. The position vector is $\mathbf{r} = x\hat{\mathbf{i}} + y\hat{\mathbf{j}}$, and the velocity is $\mathbf{v} = v_x\hat{\mathbf{i}} + v_y\hat{\mathbf{j}}$.

The numerator is the net flux of the tracer across $\partial S$:
\begin{equation}
    \oint_{\partial S} \Sigma (\mathbf{v} \cdot \hat{\mathbf{n}}) \, dl.
\end{equation}
Along the straight segment ($y=0$, $dl=dx$), the outward-pointing normal is $\hat{\mathbf{n}} = -\hat{\mathbf{j}}$. Thus, $\mathbf{v} \cdot \hat{\mathbf{n}} = -v_y$. The integral along this path is $\int_{-L}^{+L} \Sigma(x, 0) (-v_y) dx$.
Along the semicircular arc, we take the ideal limit where $L \to \infty$. For any realistic galactic disk, the surface density $\Sigma$ falls off rapidly with radius. We assume that $\Sigma \to 0$ sufficiently fast such that the integral over the arc vanishes. This is a reasonable assumption for the stellar component of disk galaxies. Consequently, the entire numerator reduces to the integral along the major axis:
\begin{equation}
    \text{Numerator} = -\int_{-\infty}^{+\infty} \Sigma(x, 0) v_y(x, 0) \, dx.
    \label{eq:classic_tw_num}
\end{equation}

Applying similar analysis to the denominator, so we have:
\begin{equation}
    \Omega_{\rm p} = \frac{-\int_{-\infty}^{+\infty} \Sigma(x, 0) v_y(x, 0) \, dx}{-\int_{-\infty}^{+\infty} \Sigma(x, 0) x \, dx} = \frac{\int \Sigma v_y \, dx}{\int \Sigma x \, dx}.
\end{equation}
By defining the luminosity-weighted average of a quantity $Q$ along the slit as $\langle Q \rangle = \int \Sigma Q \, dx / \int \Sigma \, dx$, we arrive at the celebrated Tremaine-Weinberg equation:
\begin{equation}
    \Omega_{\rm p} = \frac{\langle v_y \rangle}{\langle x \rangle}.
    \label{eq:classic_tw}
\end{equation}

This derivation highlights two key points. First, it confirms that the classic TW method is mathematically a direct consequence of the continuity equation applied to a specific, infinitely-large geometry. Second, it provides a crucial re-interpretation of what is being measured in a real-world observation. In practice, spectrographic slits are of finite length, $2L$. Therefore, the measurement does not probe the `global' pattern speed in a strict sense. Instead, according to our framework, it measures the \textit{local} pattern speed averaged over the finite area enclosed by the slit and its (implicit) closing path, weighted by the non-axisymmetric density signature as discussed in Section~\ref{sec:multiple_patterns}.

The practical power of the TW method stems from the direct relationship between the required velocity component, $v_y$, and the observable line-of-sight velocity, $v_{\rm los}$, via $v_{\rm los} = v_y \sin i$. This allows for a direct measurement of $\Omega_{\rm p}$ from long-slit spectroscopy or integral field unit (IFU) data, provided the galaxy's inclination and major axis position angle are known. The method is most robustly applied using stellar populations as the tracer, as their conservation is a very safe assumption \citep[see also][for a detailed analysis of sources of error]{2020MNRAS.491.3655G}. Applying the method to gas tracers (e.g., H$\alpha$, CO) is more problematic, as star formation and feedback can act as local sources and sinks, violating the fundamental continuity equation and leading to systematically incorrect pattern speeds \citep{2021AJ....161..185W, 2023MNRAS.524.3437B}.

Our framework also clarifies potential systematic effects. As shown in Section~\ref{sec:multiple_patterns}, if the integration path extends into regions with different dynamical components (e.g., from a bar into spiral arms), the resulting $\Omega_{\rm p}$ is a luminosity-weighted average. Many studies intentionally use slits that are much longer than the bar to establish a robust baseline. However, this means the measurement may be contaminated by the signal from the outer spiral arms. Our own tests using N-body simulations indicate that this can introduce a small systematic bias on the order of $1-2~\rm{km\,s^{-1}\,kpc^{-1}}$ ($<5\%$) on the measured bar pattern speed. This effect is subtle because the weighting (Eq.~\ref{eq:weight_definition}) is dominated by the high surface brightness of the bar, which heavily suppresses the contribution from the lower-density outer disk. Nevertheless, this ``bias" can also be interpreted as a genuine signal from the spiral arms themselves, hinting at the potential to extract more complex kinematic information if multiple patterns can be properly disentangled.

\subsection{Methods for the Milky Way Bar}
\label{sec:mw_bar}

The application of continuity-equation-based methods to our own Galaxy, where we have the advantage of full 3D phase-space information for individual stars, offers a unique testbed for pattern speed measurements. The pioneering work of \citet{2002MNRAS.334..355D} first adopted the TW method logic to the Milky Way, utilizing line-of-sight velocities to constrain the bar's pattern speed. More recently, the unparalleled astrometric precision of the \textit{Gaia} mission has enabled a more powerful formulation based on proper motions, as demonstrated by \citet{2019MNRAS.488.4552S}. Here, we demonstrate that this modern approach for the Galactic bar is also a direct and elegant consequence of our general 3D integral formulation (Eq.~\ref{eq:local_omega_p_3d}).

\begin{figure}[ht!]
    \centering
    \includegraphics[width=0.7\textwidth]{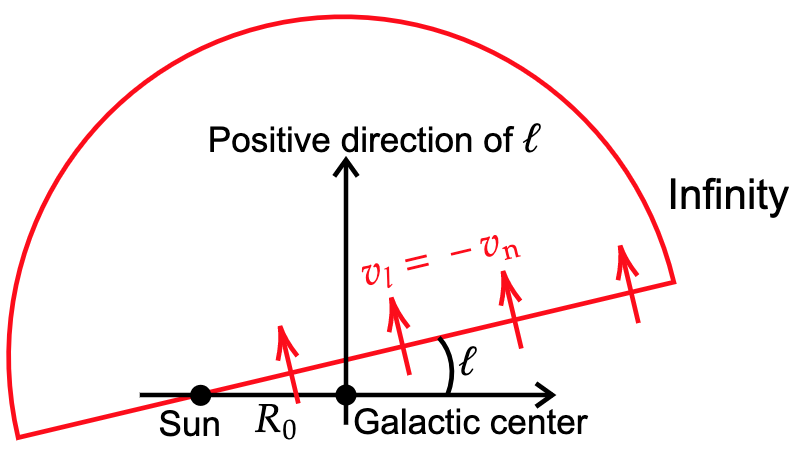} 
    \caption{Geometric schematic for deriving the \citet{2019MNRAS.488.4552S} method for the Milky Way bar from our general framework, viewed from the North Galactic Pole. The integration surface $\partial V$ is a semi-cylinder, whose flat cross-section (the straight red line) passes through the Sun and is oriented at an angle $\ell$ relative to the GC-Sun line. The red semi-circle curve represents the cylindrical wall at infinity. We apply our 3D integral formula (Eq.~\ref{eq:local_omega_p_3d}) to this volume. Because the tracer density $\rho \to 0$ at large distances, the surface integral over the curved wall vanishes. The entire contribution comes from the flat cross-section. The outward normal vector $\hat{\mathbf{n}}$ on this face points in the direction of increasing Galactic longitude. Consequently, the tracer velocity component normal to the cross-section, $\mathbf{v} \cdot \hat{\mathbf{n}}$, corresponds to the negative of longitudinal proper motion component, $-v_\ell$. This specific choice of geometry reduces our general formula to the expression used by \citet{2019MNRAS.488.4552S}.}
    \label{fig:sanders_loop}
\end{figure}

The method of \citet{2019MNRAS.488.4552S} measures the pattern speed by integrating tracer properties over a plane perpendicular to the Milky Way midplane, oriented at a specific Galactic longitude $\ell$, as viewed from the Sun. We can derive their result by choosing our integration surface $\partial V$ to be a semi-infinite cylinder, whose flat face is this very plane, and whose curved surface extends to infinity (see Figure~\ref{fig:sanders_loop}). We choose a coordinate system centered on the GC, with the Sun located at $\mathbf{R}_\odot = (-R_0, 0, 0)$.

The flat face of this cylinder is a plane containing the GC and the Galactic rotation axis, oriented at an angle $\ell$ with respect to the GC-Sun line. An outward-pointing normal vector to this plane is $\hat{\mathbf{n}} = (\sin\ell, -\cos\ell, 0)$. As with the classic 2D case, we assume the tracer density $\rho(\mathbf{x}) \to 0$ sufficiently rapidly at large distances, so the integrals over the curved part of the cylinder at infinity are zero. The measurement is therefore entirely determined by the surface integral over the flat plane.

Let us evaluate the numerator and denominator of Equation~(\ref{eq:local_omega_p_3d}) on this plane. A star on this plane, observed from the Sun with heliocentric distance $s$ and Galactic latitude $b$, has a Galactocentric position vector:
\begin{equation}
    \mathbf{r} = (s\cos b\cos\ell - R_0, s\cos b\sin\ell, s\sin b).
\end{equation}
The component of the pattern's rotational velocity field normal to the surface is:
\begin{align}
    (\hat{\mathbf{z}} \times \mathbf{r}) \cdot \hat{\mathbf{n}} &= \left( -(s\cos b\sin\ell), (s\cos b\cos\ell - R_0), 0 \right) \cdot (\sin\ell, -\cos\ell, 0) \nonumber \\
    &= -s\cos b\sin^2\ell - (s\cos b\cos\ell - R_0)\cos\ell \nonumber \\
    &= -s\cos b(\sin^2\ell + \cos^2\ell) + R_0\cos\ell \nonumber \\
    &= R_0\cos\ell - s\cos b.
\end{align}
This is the integrand for the denominator. For the numerator, we need the tracer velocity component normal to the surface, $\mathbf{v} \cdot \hat{\mathbf{n}}$. By definition, the velocity component in the direction of increasing Galactic longitude, $v_\ell$, is given by the projection of the velocity vector onto the unit vector $\hat{\mathbf{e}}_\ell = (-\sin\ell, \cos\ell, 0)$. We see immediately that $\hat{\mathbf{n}} = -\hat{\mathbf{e}}_\ell$, and therefore $\mathbf{v} \cdot \hat{\mathbf{n}} = -v_\ell$. 

Substituting these results into our 3D integral formula (Eq.~\ref{eq:local_omega_p_3d}), with the surface element being $dS = s\,ds\,db$, we get:
\begin{equation}
    \Omega_{\rm p}(\ell) = \frac{\int_{-\pi/2}^{\pi/2} db \int_{0}^{\infty} ds\, s \rho (-v_\ell)}{\int_{-\pi/2}^{\pi/2} db \int_{0}^{\infty} ds\, s \rho (R_0\cos\ell - s\cos b)}.
\end{equation}
The minus signs cancel, yielding the formulation of \citet{2019MNRAS.488.4552S}:
\begin{equation}
    \Omega_{\rm p}(\ell) = \frac{\int_{-\pi/2}^{\pi/2} db \int_{0}^{\infty} ds\, s \rho v_\ell}{\int_{-\pi/2}^{\pi/2} db \int_{0}^{\infty} ds\, s \rho (R_0\cos\ell - s\cos b)}.
    \label{eq:sanders_integral}
\end{equation}
This can be written more compactly using luminosity-weighted averages over the plane:
\begin{equation}
    \Omega_{\rm p}(\ell) = \frac{\langle v_\ell \rangle_{s,b}}{\langle R_0\cos\ell - s\cos b \rangle_{s,b}}.
    \label{eq:sanders_averaged}
\end{equation}
This derivation explicitly shows that the Sanders et al. method is a special case of our framework. The use of the longitudinal velocity component $v_\ell$ is particularly powerful as it probes the tangential motion of stars, which is strongly perturbed by the bar potential. Furthermore, it is less susceptible to contamination from the large reflex solar motion, which primarily affects line-of-sight velocity measurements.

\subsection{Methods for Local Spiral Arms in the Milky Way}
\label{sec:mw_spirals}

While the $v_\ell$-based method is effective for the large-scale bar, a different geometry may be optimal for probing the kinematics of local spiral arms. As first suggested by \citet{2002MNRAS.334..355D}, the line-of-sight velocity, $v_{\rm los}$, also contains information about the pattern speed. We propose that a method based on $v_{\rm los}$, when applied to a local volume around the Sun, is particularly well-suited to measure the pattern speeds of nearby spiral arms, which are known to induce significant radial streaming motions \citep[e.g.,][]{2011MNRAS.410.1637S}. The derivation of such a method provides another clear illustration of our framework's versatility.

\begin{figure}[ht!]
    \centering
    \includegraphics[width=0.7\textwidth]{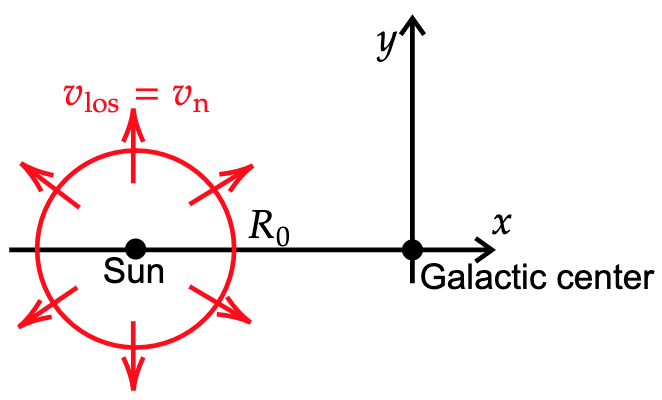}
    \caption{Geometric schematic for a proposed method to measure the pattern speeds of local spiral arms, viewed from the North Galactic Pole. The integration surface $\partial V$ is a sphere of radius $s$ centered on the Sun. The outward normal vector $\hat{\mathbf{n}}$ at any point on the sphere is, by definition, the unit vector along the line of sight. Therefore, the physical flux term in our formalism, $\rho (\mathbf{v} \cdot \hat{\mathbf{n}})$, directly corresponds to the flux measured via the line-of-sight velocity, $\rho v_{\rm los}$. By integrating over this spherical surface, our general formula (Eq.~\ref{eq:local_omega_p_3d}) yields a direct relationship between the pattern speed $\Omega_{\rm p}$ and the observed distribution of line-of-sight velocities and densities on the sky, making it a powerful tool for analyzing data from large-scale stellar surveys.}
    \label{fig:sphere_loop}
\end{figure}

To isolate local structures, the natural integration surface is a sphere of radius $s$ centered on the Sun, as depicted in Figure~\ref{fig:sphere_loop}. This geometry is ideal for modern spectroscopic surveys that provide distances and line-of-sight velocities for nearly complete samples of stars within a given volume.

We again start from Equation~(\ref{eq:local_omega_p_3d}) and evaluate the integrals over this spherical surface. The outward normal vector $\hat{\mathbf{n}}$ is simply the unit vector in the heliocentric radial direction:
\begin{equation}
    \hat{\mathbf{n}} = (\cos b\cos\ell, \cos b\sin\ell, \sin b).
\end{equation}
The Galactocentric position vector $\mathbf{r}$ of a star on the sphere is the same as before: $\mathbf{r} = (s\cos b\cos\ell - R_0, s\cos b\sin\ell, s\sin b)$. The pattern's kinematic flux component is then:
\begin{align}
    (\hat{\mathbf{z}} \times \mathbf{r}) \cdot \hat{\mathbf{n}} &= \left( -(s\cos b\sin\ell), (s\cos b\cos\ell - R_0), 0 \right) \cdot (\cos b\cos\ell, \cos b\sin\ell, \sin b) \nonumber \\
    &= -s\cos^2 b\sin\ell\cos\ell + (s\cos b\cos\ell - R_0)\cos b\sin\ell \nonumber \\
    &= -s\cos^2 b\sin\ell\cos\ell + s\cos^2 b\sin\ell\cos\ell - R_0\cos b\sin\ell \nonumber \\
    &= -R_0\cos b\sin\ell.
\end{align}
The numerator's velocity term, $\mathbf{v} \cdot \hat{\mathbf{n}}$, is by definition the stellar line-of-sight velocity, $v_{\rm los}$, measured by an observer at the Sun. The surface element on the sphere is $dS = s^2 \cos b\,d\ell\,db$.

Substituting these into Equation~(\ref{eq:local_omega_p_3d}) gives the pattern speed as a function of the integration sphere's radius $s$:
\begin{equation}
    \Omega_{\rm p}(s) = \frac{\int_{0}^{2\pi} d\ell \int_{-\pi/2}^{\pi/2} db\, s^2\cos b\, \rho v_{\rm los}}{\int_{0}^{2\pi} d\ell \int_{-\pi/2}^{\pi/2} db\, s^2\cos b\, \rho (-R_0\cos b\sin\ell)}.
\end{equation}
In the compact, luminosity-averaged form over the celestial sphere at distance $s$, this becomes:
\begin{equation}
    \Omega_{\rm p}(s) = \frac{\langle v_{\rm los} \rangle_{\ell,b}}{-R_0\langle \cos b\sin\ell \rangle_{\ell,b}}.
    \label{eq:vlos_averaged}
\end{equation}

It is instructive to note that this shell-based result generalizes the seminal method of \citet{2002MNRAS.334..355D}. In the pre-\textit{Gaia} era, precise distances to individual stars were not available to isolate thin shells, necessitating an integration over the entire observable cone. If we integrate the numerator and denominator of the expanded form above over all distances $s$ (weighted by a detection probability $f(s)$), we effectively sum the contributions of concentric spheres. This operation recovers the exact volume-integrated expression derived by \citet{2002MNRAS.334..355D} for the quantity $\Omega_{\rm p} R_0$ (their Eqs.~8 and 9):
\begin{equation}
    \Omega_{\rm p} R_0 = \frac{\int ds \int d\ell \int db \, (s^2 f(s) \rho) \, v_{\rm los} \cos b}{-\int ds \int d\ell \int db \, (s^2 f(s) \rho) \, \cos^2 b \sin \ell}.
\end{equation}
Thus, the D02 method represents the volume-integrated limit of our framework. Our differential formulation (Eq.~\ref{eq:vlos_averaged}), enabled by modern data, allows for a tomographic approach: it establishes a direct, model-independent link between the \textit{local} pattern speed and angular variations in tracer density and line-of-sight velocity at any specific distance. 

If spiral arms generate strong, systematic $v_{\rm los}$ signatures as a function of Galactic longitude, this method is an ideal tool for measuring their pattern speeds. A full application of this technique using a volume-complete sample of young stars will be presented in a subsequent paper. Its derivation here completes our demonstration that continuity-equation methods for both the bar and spiral arms in the Milky Way are unified under our general integral framework.

\subsection{The Radial TW Method and its Geometric Approximation}
\label{sec:radial_tw}

\begin{figure}[ht!]
    \centering
    \includegraphics[width=0.9\textwidth]{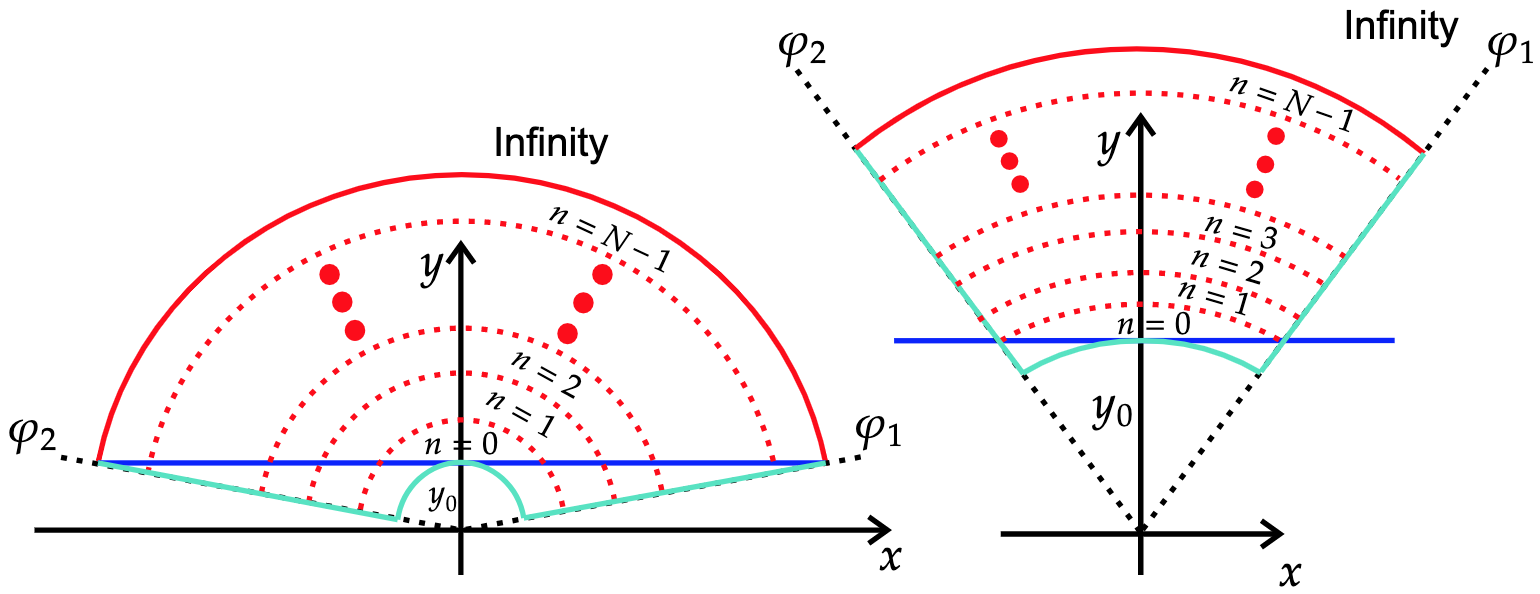}
    \caption{
    Geometric comparison between the traditional radial TW formulation and the exact flux balance required by the continuity equation. 
    \textit{Left panel:} The geometry near the galaxy center (small projected distance $y_0$).
    \textit{Right panel:} The geometry at a larger projected distance (large $y_0$), where the deviation becomes significant.
    In both panels, the cyan curve represents the theoretical inner boundary of the integration domain implied by the LHS of the traditional radial TW equation (Eq.~\ref{eq:radial_tw_original}), while the solid blue line indicates the observational slit at a constant vertical position $y=y_0$ (the RHS of Eq.~\ref{eq:radial_tw_original}). Note how the geometric discrepancy between the curved boundary and the straight slit increases with $y_0$.
    }
    \label{fig:radial_tw_approx}
\end{figure}

While the classic TW method (Section~\ref{sec:classic_tw}) and its variants for the Milky Way (Sections~\ref{sec:mw_bar} and \ref{sec:mw_spirals}) are powerful, they are designed to measure a single, average pattern speed for the region enclosed by the integration loop. However, galaxies can host dynamically distinct components, such as a bar and spiral arms with different rotational speeds, or transient spiral structures whose pattern speeds vary with radius \citep[e.g.,][]{2011MNRAS.410.1637S}. To address this complexity, a ``general" or ``radial" TW method was developed to solve for a radially dependent pattern speed, $\Omega_{\rm p}(r)$. This approach was pioneered by \citet{1994A&A...285..801E} and later refined and applied in works by \citet{2008ApJ...676..899M}, \citet{2009ApJ...702..277M}, and \citet{2016ApJ...826....2S}.

While the classic TW method (Section~\ref{sec:classic_tw}) and its variants for the Milky Way (Sections~\ref{sec:mw_bar} and \ref{sec:mw_spirals}) are powerful, they are designed to measure a single, average pattern speed for the region enclosed by the integration loop. However, galaxies can host dynamically distinct components, such as a bar and spiral arms with different rotational speeds, or transient spiral structures whose pattern speeds vary with radius \citep[e.g.,][]{2011MNRAS.410.1637S}. To address this complexity, a ``general" or ``radial" TW method was developed to solve for a radially dependent pattern speed, $\Omega_{\rm p}(r)$. This approach was pioneered by \citet{1994A&A...285..801E} and later refined and applied in works by \citet{2008ApJ...676..899M}, \citet{2009ApJ...702..277M}, and \citet{2016ApJ...826....2S}. The derivations in this section are intricate. To maintain the integrity of the paper while enhancing readability for the primary text, we have relocated these detailed derivations to the appendix.

These works begin with a continuity equation that, when integrated over a domain covering half the galaxy ($y > 0$), yields an integral equation for the pattern speed $\Omega_{\rm p}(r)$. For a series of slits aligned parallel to the galaxy's major axis, each at a different projected distance $y$ from the center, the equation takes the form:
\begin{equation}
    \int_{y}^{\infty} \left\{ \left[ \Sigma(\sqrt{r^2-y^2}, y) - \Sigma(-\sqrt{r^2-y^2}, y) \right] \frac{r}{ \sqrt{r^2-y^2} } \right\} \Omega_{\rm p}(r) dr = \int_{-\infty}^{\infty} \Sigma v_y dx.
    \label{eq:radial_tw_original}
\end{equation}
In this formulation, the right-hand side is the total observed mass flux across a line of constant $y$, and the left-hand side is an integral transform of the unknown function $\Omega_{\rm p}(r)$. By observing the flux for many different values of $y$, one can in principle deconvolve this equation—a Fredholm integral equation of the first kind—to recover the pattern speed profile. While powerful in concept, this method contains a subtle but important geometric approximation that our framework can make explicit.

We can re-derive the essence of this method by starting from our analysis of multiple pattern speeds in Section~\ref{sec:multiple_patterns}. Consider the summed flux balance for a stack of thin, contiguous annular sectors, as expressed in Equation~(\ref{eq:summed_balance}). We analyze a region extending from an inner radius $r=r_0$ to infinity, bounded by azimuths $\varphi_1$ and $\varphi_2$. In the continuum limit where the width of each sub-annulus approaches zero, the sum on the left-hand side of Equation~(\ref{eq:summed_balance}) becomes a definite integral. The total flux on the right-hand side, $\mathcal{F}_{\rm total}$, consists of contributions from the four boundaries of the full sector. Assuming the tracer density and flux vanish at infinity, the only non-zero contributions are from the two radial lines and the inner arc at $r=r_0$. The exact balance equation is therefore:
\begin{equation}
    \sum_n \left\{\left[\Sigma(r_n, \varphi_1)-\Sigma(r_n, \varphi_2)\right] r_n \delta r\right\} \Omega_p(r_n) = \\\left.\left(\int_{r_0}^{\infty} \Sigma v_\varphi dr\right)\right|_{\varphi_1}^{\varphi_2} + \left.\left(\int_{\varphi_1}^{\varphi_2} \Sigma v_r rd\varphi\right)\right|_{r_0}^{\infty}.
    \label{eq:exact_radial_balance}
\end{equation}

The connection to, and the approximation within, the traditional radial TW method becomes clear when we choose a specific geometry, as illustrated in Figure~\ref{fig:radial_tw_approx}. Let the integration domain be the upper half of the galactic disk, symmetric about the $y$-axis. Here, the radial boundaries correspond to azimuths $\varphi_1$ and $\varphi_2$ that approach $0$ and $\pi$, and the inner boundary is a semicircle of radius $r_0 = y_0$. In this geometry, the left-hand side integral in Equation~(\ref{eq:exact_radial_balance}) becomes equivalent to the left-hand side of Equation~(\ref{eq:radial_tw_original}) after a change of coordinates.

However, the right-hand sides differ critically. The traditional method's right-hand side, $\int \Sigma v_y dx$, represents the total mass flux across a \textit{straight line} at a constant height $y_0$ above the major axis (the blue line in Figure~\ref{fig:radial_tw_approx}). In contrast, our exact derivation shows that for the continuity equation to hold over the chosen annular sector, the flux that truly balances the pattern's rotation is the flux across the boundaries of that sector. This includes the flux across the \textit{curved arc} at $r=r_0$ (the cyan line in Figure~\ref{fig:radial_tw_approx}).

Therefore, the traditional radial TW method makes an implicit approximation: it substitutes the easily observable flux integral along a straight Cartesian slit for the physically required flux integral along a curved polar-coordinate boundary. As Figure~\ref{fig:radial_tw_approx} illustrates, this approximation is reasonable only when the integration starts very close to the galactic center (small $y_0$), where the arc is nearly flat. For slits placed at larger distances from the center, the geometric discrepancy between the straight line and the circular arc becomes severe. The path length, the local normal vector $\hat{\mathbf{n}}$, and the tracer properties ($\Sigma, \mathbf{v}$) all differ between the two paths, leading to a systematic error in the measured flux balance.

This approximation was a practical necessity, driven by the operational ease of obtaining and analyzing data from long slits. However, our framework reveals that it is the source of a potentially significant geometric error. This insight directly motivates the need for a revised formulation that adheres to an exact geometry, which we will develop in the following section. By choosing an integration loop for which all boundary integrals are observationally tractable without approximation, we can construct a more robust method for measuring radially varying pattern speeds.

\subsection{Our Matrix Formulation for Radially Varying Pattern Speeds}
\label{sec:refined_radial_tw}

\begin{figure}[ht!]
    \centering
    \includegraphics[width=0.8\textwidth]{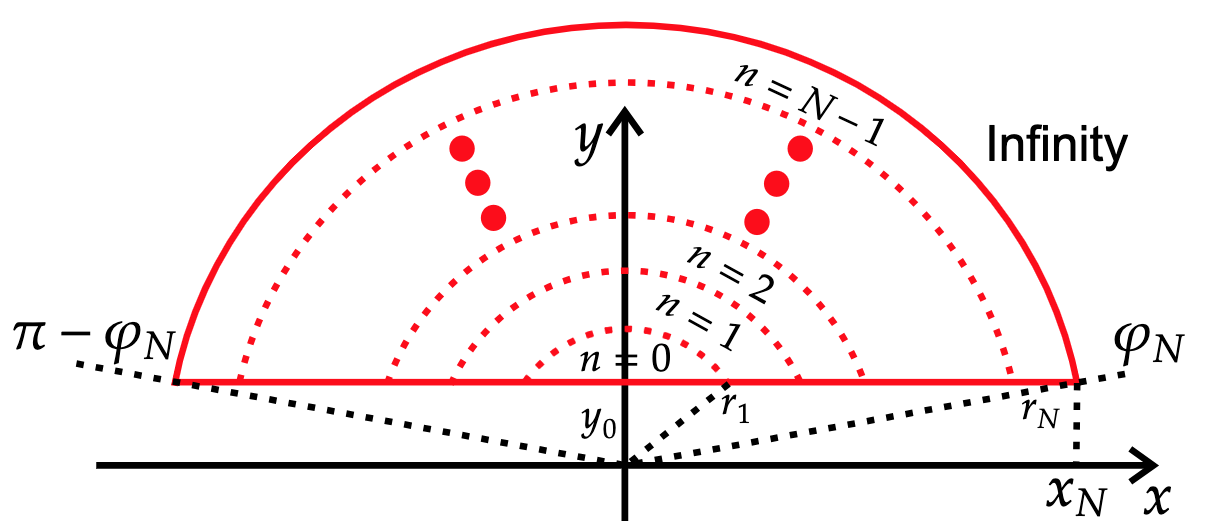}
    \caption{
    Schematic of the geometry for geometrically exact formulation radial TW method of a single slit when $m=0$. The integration loop (solid red line) consists of a straight line segment at $y=y_0$, analogous to an observational slit, and a closing semicircle at a large radius $r_N \to \infty$. The enclosed area is subdivided by $N$ concentric circles (dotted red lines) at radii $r_n$ with all their centers at the origin, defining a series of radial bins where the pattern speed $\Omega_{\rm p}(r_n)$ is assumed to be constant. Unlike traditional formulations, this method does not approximate curved boundaries with straight ones. All flux and moment integrals are calculated on their exact geometric paths. This approach leverages the conceptual clarity of our local pattern speed framework (Section~\ref{sec:framework}): by choosing an integration loop that is both physically exact and observationally practical, we can build a exact method for measuring radially varying pattern speeds.
    }
    \label{fig:refined_tw_geo}
\end{figure}

\begin{figure}[ht!]
    \centering
    \includegraphics[width=0.8\textwidth]{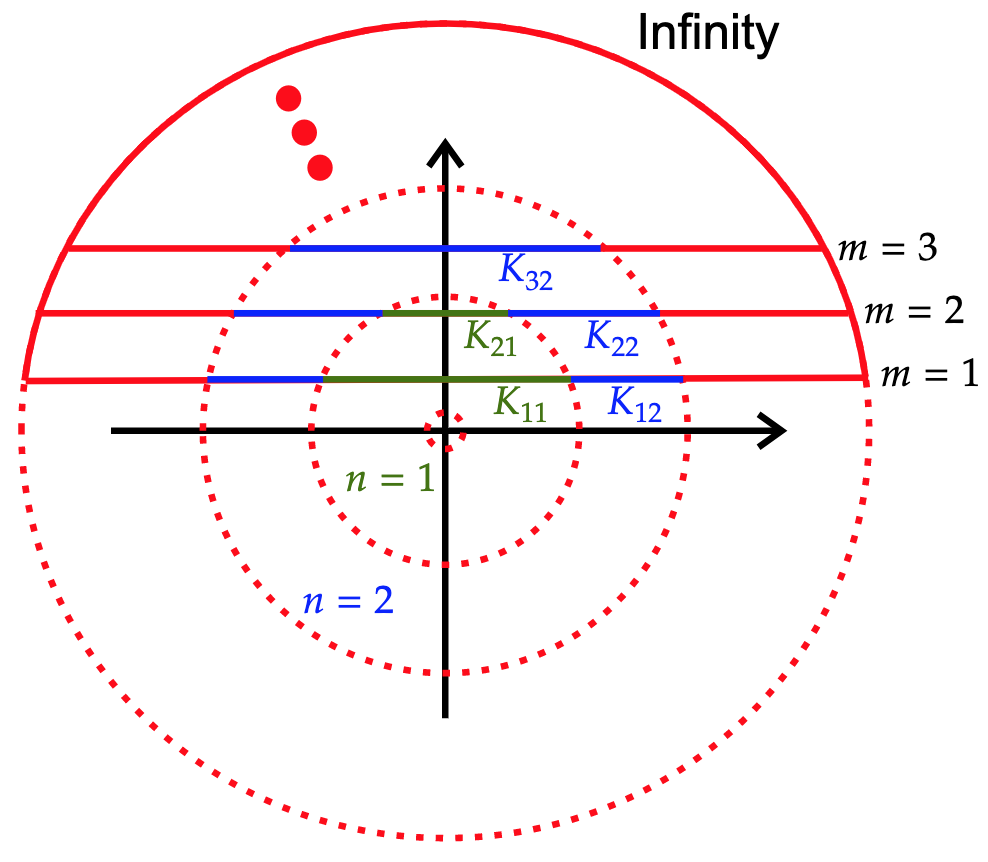}
    \caption{
    Visualization of the matrix elements for the linear system $\mathbf{K}\boldsymbol{\Omega} = \mathbf{W}$ (Eq.~\ref{eq:matrix_equation}). We use multiple slits ($m=1, 2, 3, \ldots$) positioned at different projected heights $y_m$ to probe a disk subdivided into concentric radial annuli ($n=1, 2, \ldots$). The matrix element $\mathbf{K}_{mn}$ (represented by the colored line segments) is the luminosity-weighted moment integral $\int \Sigma x \, dx$ calculated along the portions of the $m$-th slit that pass through the $n$-th annulus. This element quantifies how the pattern in annulus $n$ (with speed $\Omega_n$) influences the net mass flux measured across the entirety of slit $m$. By combining measurements from multiple slits that sample the disk differently, we can construct an overdetermined system and solve for the detailed radial profile of the pattern speed $\boldsymbol{\Omega}$.
    }
    \label{fig:matrix_setup}
\end{figure}

The analysis in Section~\ref{sec:radial_tw} revealed that traditional formulations of the radial Tremaine-Weinberg method rely on a geometric approximation: equating the tracer flux across a physically curved boundary with the observationally uniform flux across a straight Cartesian slit. While valid near the galactic center, this approximation introduces systematic errors at larger radii. Here, we present a geometrically exact formulation using our integral framework.

We propose a refined integration path composed of circular arcs, as illustrated in Figure~\ref{fig:refined_tw_geo} (see panel b). By centering all arcs on the coordinate origin, we avoid any geometric approximations. This approach naturally degrades to the standard TW method for $N=1$, demonstrating the flexibility of path selection based on local pattern speed requirements.

We divide the galaxy into $N$ concentric annuli. Within the $n$-th annulus bounded by radii $r_n$ and $r_{n+1}$, the pattern speed $\Omega_{\rm p}(r_n)$ is well-defined. We apply Equation~\ref{eq:general_loop_integral} to the area enclosed by the intersection of a horizontal slit at $y=y_m$ and the $n$-th annulus, where the boundary $\mathcal{L}_n$ comprises two straight horizontal segments and two circular arcs at $r_n$ and $r_{n+1}$. Let the straight line segments extend over $x \in [-x_{n+1}, -x_n] \cup [x_n, x_{n+1}]$. We have:
\begin{equation}
    \Omega_{\rm p}(r_n) \left(\int_{-x_{n+1}}^{-x_n}+\int_{x_n}^{x_{n+1}}\right) \Sigma x \, dx = R_n - R_{n+1} + \left(\int_{-x_{n+1}}^{-x_n}+\int_{x_n}^{x_{n+1}}\right) \Sigma v_y \, dx,
    \label{eq:exact_slit_equation}
\end{equation}
where the flux contributions across the circular boundaries, representing radial transport, are defined as:
\begin{equation}
    R_n \equiv \left( \int_{\phi_n}^{\pi-\phi_n} \Sigma v_r r \, d\phi \right)_{r=r_n}.
\end{equation}

We then sum Equation~(\ref{eq:exact_slit_equation}) over all $N$ annuli, from the center to a sufficiently large radius $r_N$ where density vanishes ($R_N \to 0$, $R_0 = 0$). The internal radial flux terms $R_n$ cancel out telescopically, yielding the exact integral equation:
\begin{equation}
    \sum_{n=0}^{N-1} \Omega_{\rm p}(r_n) \left[\left(\int_{-x_{n+1}}^{-x_n}+\int_{x_n}^{x_{n+1}}\right) \Sigma x \, dx \right]_{y=y_m} = \int_{-x_N}^{x_N} \Sigma v_y \, dx \bigg|_{y=y_m}.
\end{equation}
This summation eliminates the unobservable radial fluxes, connecting the weighted sum of pattern speeds directly to the observable line-of-sight velocity integral along the slit.

By collecting data from $M$ slits, we construct a system of linear equations. This system can be expressed in matrix form, as illustrated in Figure~\ref{fig:matrix_setup}:
\begin{equation}
    \mathbf{K} \boldsymbol{\Omega} = \mathbf{W},
    \label{eq:matrix_equation}
\end{equation}
where $\boldsymbol{\Omega}$ is the $N \times 1$ vector of unknown pattern speeds, $\mathbf{W}$ is the $M \times 1$ vector of observed fluxes, and $\mathbf{K}$ is the $M \times N$ kernel matrix that couples them. Their elements are defined as follows:
\begin{gather}
    \boldsymbol{\Omega}_n = \Omega_{\rm p}(r_n), \\
    \mathbf{W}_m = \left( \int_{-x_{m,N}}^{x_{m,N}} \Sigma v_y \, dx \right)_{y=y_m}, \\
    \mathbf{K}_{mn} = \left( \int_{-x_{m,n+1}}^{-x_{m,n}} + \int_{x_{m,n}}^{x_{m,n+1}} \right) \Sigma x \, dx \bigg|_{y=y_m}. \label{eq:matrix_elements}
\end{gather}
The term $\mathbf{K}_{mn}$ represents the luminosity-weighted first moment of the coordinate $x$ along the segments of the $m$-th slit that lie within the $n$-th radial annulus. This matrix element quantifies the contribution of the pattern in annulus $n$ to the total expected flux across slit $m$.

In practice, to obtain a well-constrained solution for $\boldsymbol{\Omega}$, one typically uses more slits than radial bins ($M > N$), resulting in an overdetermined system. The optimal solution in a least-squares sense is found by solving the normal equations, or, more robustly, by employing the pseudoinverse of the kernel matrix, $\mathbf{K}^\dagger$:
\begin{equation}
    \boldsymbol{\Omega} = \mathbf{K}^\dagger \mathbf{W}.
    \label{eq:svd_solution}
\end{equation}
The pseudoinverse is readily calculated via Singular Value Decomposition (SVD), a powerful numerical technique that decomposes the kernel matrix as $\mathbf{K} = \mathbf{U} \mathbf{S} \mathbf{V}^T$. Here, $\mathbf{U}$ and $\mathbf{V}$ are orthogonal matrices, and $\mathbf{S}$ is a diagonal matrix of singular values. The pseudoinverse is then constructed as $\mathbf{K}^\dagger = \mathbf{V} \mathbf{S}^\dagger \mathbf{U}^T$, where $\mathbf{S}^\dagger$ is formed by taking the reciprocal of the non-zero singular values in $\mathbf{S}$. SVD provides a stable way to solve the system, even in the presence of degeneracies or near-zero singular values, which correspond to modes of the pattern speed profile that are poorly constrained by the data.

A remaining practical challenge is the choice of radial bins, i.e., the placement of the radii $r_n$. Previous studies have explored different approaches. For example, one can employ regularization techniques to enforce smoothness on the solution $\boldsymbol{\Omega}$, effectively penalizing large differences between adjacent $\Omega_n$ values \citep[e.g.,][]{2008ApJ...676..899M}. While this improves the signal-to-noise ratio, it can introduce a significant bias, as it tends to smooth over genuine sharp transitions in pattern speed, such as those expected between a bar and a separate spiral pattern. An alternative, data-driven approach is to use model selection criteria, such as the Bayesian Information Criterion (BIC), to recursively determine the optimal number and location of the breakpoints $r_n$ that are statistically justified by the data \citep{2016ApJ...826....2S}. A full exploration of these data analysis techniques is beyond the scope of this paper, but our derivation of the exact linear system in Equation~(\ref{eq:matrix_equation}) provides the necessary and robust foundation for such advanced analyses.

\section{Validation on a Prototype Barred Galaxy}
\label{sec:validation}

To validate the robustness of the local pattern speed framework derived in Section~\ref{sec:framework}, we apply it to a barred galaxy selected from the TNG50-1 cosmological magneto-hydrodynamical simulation \citep{2019MNRAS.490.3196P, 2019MNRAS.490.3234N}. The high mass resolution of TNG50 allows us to probe the internal kinematics of galactic structures. We focus on Subhalo ID 585282 at Snapshot 99 ($z=0$), a Milky Way-mass galaxy hosting a strong bar (For more details of TNG50, see our Section~\ref{sec:diversity_profile}). In the following section, we will extend the analysis to a diverse sample of limited galaxies to explore the variety of dynamical states naturally occurring in a cosmological context.

\subsection{Radial Recovery of Pattern Speed}
\label{sec:validation_radial}

To validate our method, we first apply Equation~(\ref{eq:local_pattern_speed}) to a face-on disk galaxy (TNG50-1 ID 585282) to investigate whether it can correctly recover the radial profile of the pattern speed, distinguishing the coherently rotating bar from the differential rotation of the disk.

\subsubsection{Methodology}

We aligned the galaxy such that the principal axis of the bar coincides with the $x$-axis. The stellar particles were binned using Voronoi Tessellation to ensure a target S/N~$>10$ per bin. The Voronoi-binned data were subjected to Fourier decomposition and reconstruction, retaining terms up to the sixteenth harmonic ($m=16$). This decomposition facilitated the subsequent interpolation process. This procedure smooths local Poisson noise while preserving non-axisymmetric features essential for the pattern speed determination.

For the calculation of the local pattern speed $\Omega_{\rm p}(R)$, we utilized a series of overlapping annular sectors (Equation~\ref{eq:local_pattern_speed}) with a radial width $\Delta R = 0.5$ kpc and an azimuthal opening angle $\Delta \varphi = 30^\circ$. For each radial bin, we performed a linear fit of the numerator (flux term) against the denominator (density contrast moment) across all azimuthal sectors to determine the best-fitting $\Omega_{\rm p}$ and its uncertainty.

We determine the bar extent, $R_{\rm bar}$, by analyzing the $m=2$ Fourier component of the surface density, following the methodology implemented in the Dehnen's code \citep{2023MNRAS.518.2712D} and the criteria outlined in \citet{2022MNRAS.512.5339R}. The determination of the bar radius is based on two complementary criteria:

\begin{enumerate}
    \item \textbf{Bar Strength:} We search for the peak of the bisymmetric Fourier amplitude, $A_2 \equiv \Sigma_2/\Sigma_0$, within a fiducial range ($0.3 \le R \le 4.5$ kpc). We require a peak amplitude $A_{2,\rm peak} > 0.2$ to classify the structure as a bar. The bar region is then extended radially from the peak outwards and inwards as long as $A_2$ remains above $0.5 \times A_{2,\rm peak}$ and does not fall below a floor of 0.15.
    \item \textbf{Phase Coherence:} We further constrain the bar region by requiring the phase angle of the $m=2$ mode, $\phi_2(R)$, to be nearly constant. The total variation of the phase angle within the bar region, $|\phi_{2, \rm max} - \phi_{2, \rm min}|$, must not exceed $10^\circ$.
\end{enumerate}

To assess the reliability of our local pattern speed measurement in regions dominated by different non-axisymmetric features, we decompose the surface density into Fourier modes up to $m=16$. We define a cumulative non-axisymmetry strength parameter, $f_{\rm sum} = \sum_{m=1}^{16} A_m$. Since the derivation of a pattern speed is only physically meaningful where a coherent non-axisymmetric perturbation exists, we apply a validity threshold of $f_{\rm sum} > 0.3$. Measurements in regions falling below this threshold are considered noise-dominated and are visually de-emphasized in our analysis.

\subsubsection{Results}

\begin{figure*}[ht!]
    \centering
    \includegraphics[width=\textwidth]{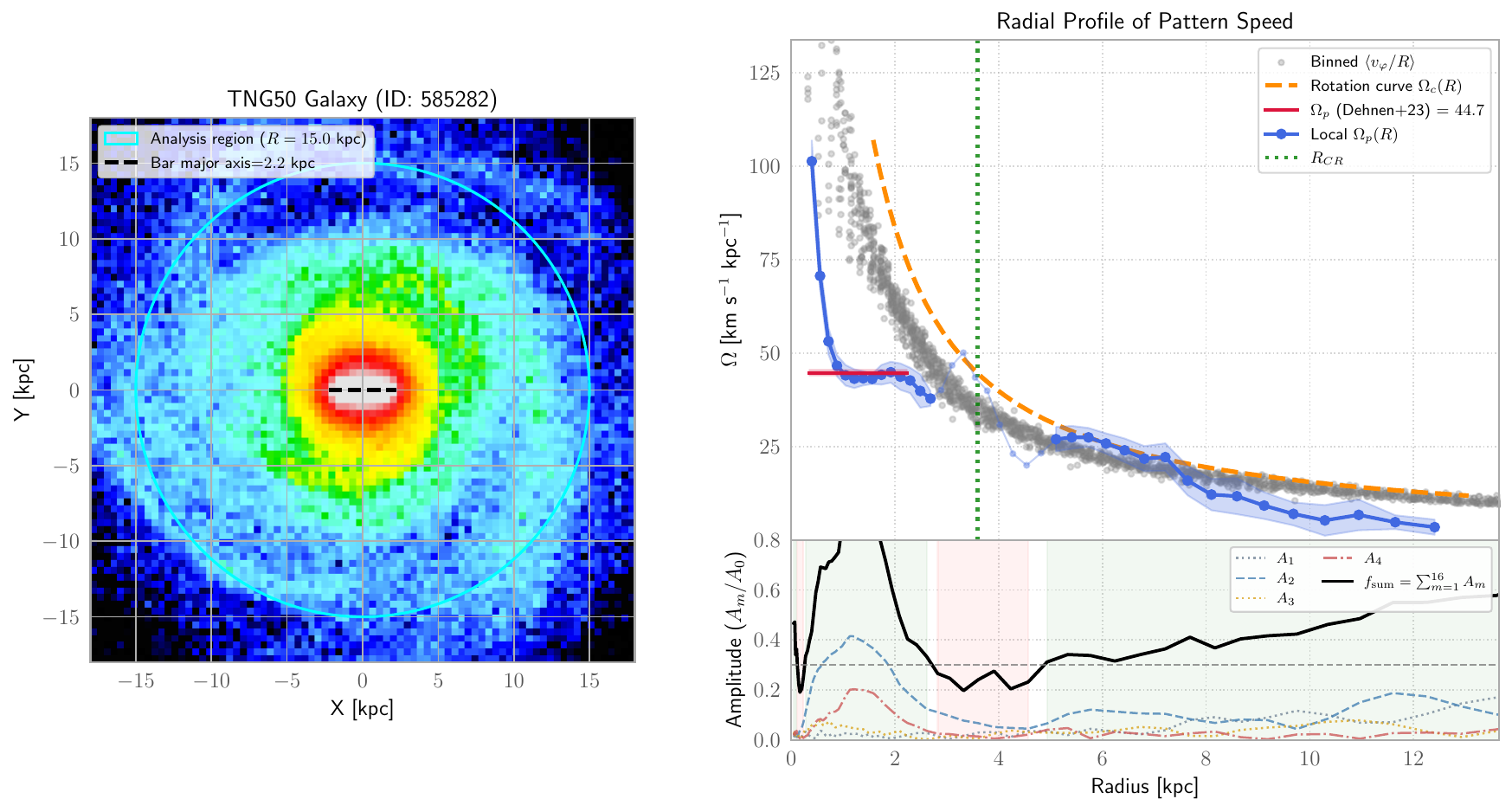}
    \caption{An example of local pattern speed on TNG50-1 galaxy ID 585282. 
    \textbf{Left Panel:} Face-on stellar surface density map. The analysis region extends to $R=15.0$ kpc (cyan circle). The bar is aligned along the X-axis, with its length indicated by the black dashed line ($R_{\rm bar} \approx 2.2$ kpc).
    \textbf{Right Panels:} Radial analysis of the pattern speed and structural modes. 
    \textbf{Top Subplot:} The blue solid line with markers represents the local pattern speed, $\Omega_{\rm p}(R)$. Regions where the non-axisymmtry signal is weak ($f_{\rm sum} < 0.3$) are shown with reduced opacity. Shaded regions indicate $1\sigma$ uncertainties. Grey dots represent the binned tangential verlocity $\langle v_\varphi/R \rangle$. The red horizontal line indicates the bar pattern speed ($\Omega_{p} \approx 44.7 \, \mathrm{km \, s^{-1} \, kpc^{-1}}$) calculated using the method of \citep{2023MNRAS.518.2712D} over the identified bar region. The orange dashed line shows the circular frequency $\Omega_c(R)$, and the vertical green dotted line marks the corotation radius ($R_{\rm CR}$). 
    \textbf{Bottom Subplot:} Radial profiles of the relative Fourier amplitudes $A_m/A_0$ for modes $m=1$ to $m=4$. The grey horizontal dashed line marks the threshold ($f_{\rm sum}=0.3$) used to filter reliable pattern speed measurements in the upper panel.}
    \label{fig:radial_validation}
\end{figure*}

The results are presented in Figure~\ref{fig:radial_validation}. The right panels illustrate the capability of our method to resolve the radial structure of the pattern speed. In the top subplot, the local pattern speed profile (blue curve) exhibits a distinct flat plateau within the inner region ($R \lesssim 2.5$ kpc). The value of this plateau aligns remarkably well with the bar pattern speed ($\Omega_{p} \approx 44.7 \, \mathrm{km \, s^{-1} \, kpc^{-1}}$) calculated using the method of \citep{2023MNRAS.518.2712D} over the identified bar region. This confirms that our local formulation accurately recovers the rigid rotation of the bar.

The bottom subplot provides insight into the structural composition of the galaxy. The bar region is dominated by the $m=2$ mode (dashed blue line), typical of bisymmetric bars. However, we also observe contributions from other modes: $m=1$ (lopsidedness), $m=3$ (triangular distortions), and $m=4$ (boxiness). The total non-axisymmetry $f_{\rm sum}$ (black line) is used to filter reliable local pattern speed measurements.

It is important to note that our method is not limited to measuring the pattern speed of $m=2$ structures (bars or grand-design spirals). As indicated by the variations in $\Omega_{\rm p}(R)$ beyond the corotation radius ($R_{\rm CR} \approx 3.5$ kpc), the method is sensitive to other dynamical features where $f_{\rm sum}$ remains significant. The effective detection of these diverse, higher-order patterns is a rigorous test of our methodology, which will be explored in greater detail in Section~\ref{sec:diversity_profile}. 

\subsection{Validation in a 3D ``Milky Way'' Configuration}
\label{sec:validation_mw}

As a second test, we validate the 3D integral formulation (Section~\ref{sec:mw_bar}) in a configuration designed to mimic \textit{Gaia}-like observations of the Milky Way as \citet{2019MNRAS.488.4552S} did. This test assesses the method's performance under ultra ideal observational constraints, including a fixed viewing angle and heliocentric coordinate geometry.

\subsubsection{Methodology}

We placed a mock observer (``Sun") in the galactic plane ($z=0$) at a distance $R_0 = 8.1$ kpc from the center. The galaxy was rotated by $\phi_{\rm bar} = -27^\circ$ relative to the Sun-Galactic Center line to match estimates of the Milky Way bar orientation \citep[e.g.,][]{2020RAA....20..159S}. We converted the simulation data into heliocentric coordinates: distance $s$, Galactic longitude $\ell$, and latitude $b$.

To ensure the validity of the tracer conservation assumption and mimic observational selection functions, we applied the following geometric cuts:
\begin{itemize}
    \item \textbf{Latitude:} $|b| < 10^\circ$, isolating the disk/bar population.
    \item \textbf{Distance:} $1.0 < s < 15.0$ kpc. The lower cut avoids singularities near the observer, while the upper cut reduces contamination from the stellar halo.
\end{itemize}

We implemented the discrete version of the 3D pattern speed estimator (Equation~\ref{eq:local_omega_p_3d}). For a bin centered at Galactic longitude $\ell_c$, the pattern speed is given by:
\begin{equation}
    \Omega_{\rm p}(\ell_c) = \frac{\sum_{i} m_i (\mathbf{v}_i \cdot \hat{\mathbf{n}}_{\ell}) \mathcal{W}_i}{\sum_{i} m_i [(\hat{\mathbf{z}} \times \mathbf{r}_i) \cdot \hat{\mathbf{n}}_{\ell}] \mathcal{W}_i},
    \label{eq:discrete_estimator}
\end{equation}
where the sum runs over all particles within the bin $|\ell_i - \ell_c| < \Delta \ell/2=1^\circ$ (under ideal mathematical conditions, we should choose an infinitesimally thin cross-section where $\Delta \ell \to 0$). Here, $\hat{\mathbf{n}}_{\ell} = (\sin \ell_c, -\cos \ell_c, 0)$ is the normal vector to the integration plane. The term $\mathcal{W}_i$ is a Jacobian weight required to convert the volume sum (over the pyramidal volume defined by the solid angle of the bin) into an approximation of the surface integral demanded by Equation~(\ref{eq:local_omega_p_3d}):
\begin{equation}
    \mathcal{W}_i = \frac{1}{s_i \cos b_i}.
\end{equation}
This weight accounts for the fact that the volume of a solid angle element scales as $s^2$, while the required integration is over an area element $dS = s\, ds\, db$.

\subsubsection{Results}

\begin{figure*}[ht!]
    \centering
    \includegraphics[width=\textwidth]{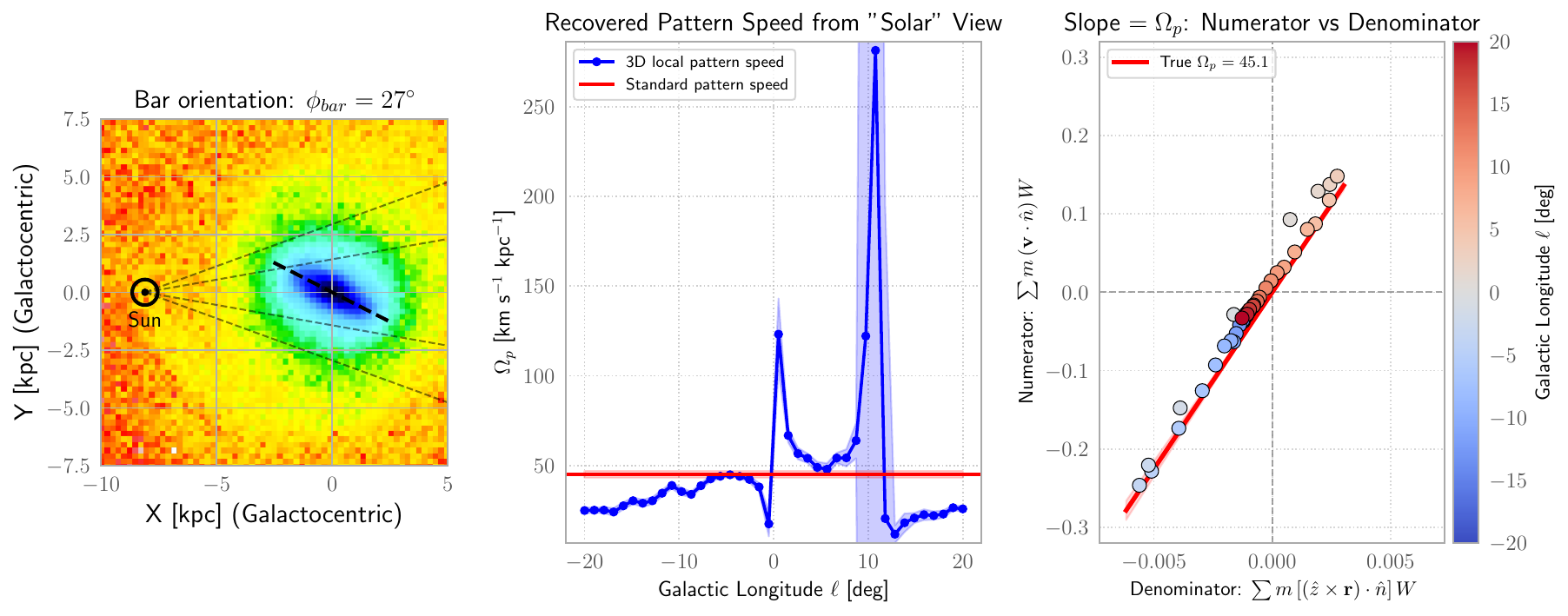}
    \caption{Validation of the 3D pattern speed measurement using a TNG50 Milky Way analogue.
    \textbf{Left panel (a):} Face-on density map rotated to a bar angle of $\phi_{\rm bar} = 27^\circ$. The ``Sun" is located at $X=-8.1$ kpc (black circle). Four thin dashed lines extending from the Sun indicate the lines of sight for Galactic longitudes $\ell = \pm 10^\circ$ and $\ell = \pm 20^\circ$. The thick black dashed line marks the extent and orientation of the stellar bar.
    \textbf{Middle panel (b):} The recovered pattern speed $\Omega_{\rm p}$ as a function of Galactic longitude $\ell$. Blue points show the bin-by-bin calculation. While the method accurately recovers the true value (red line, $\Omega_{p,\rm true} \approx 45$ km s$^{-1}$ kpc$^{-1}$) for $|\ell| \sim 5^\circ$, it exhibits divergences near $\ell \approx 0^\circ$ and $\ell \approx 12^\circ$. These features arise where line-of-sight symmetries cause the denominator of Equation~(\ref{eq:discrete_estimator}) to approach zero.
    \textbf{Right panel (c):} A robust diagnostic plot: the TW numerator term versus the denominator term for all longitude bins. Points are color-coded by $\ell$. The strict linear correlation demonstrates that the kinematic signal is dominated by a single rotating pattern. The slope of the red regression line yields the true pattern speed, effectively weighting the high-signal regions and bypassing the local divergences seen in the middle panel.}
    \label{fig:mw_validation}
\end{figure*}

Figure~\ref{fig:mw_validation} summarizes the performance of the method. Panel (a) illustrates the experimental setup, displaying the face-on surface density of the TNG50 Milky Way analogue. The galaxy is rotated such that the bar makes an angle of $\phi_{\rm bar} = 27^\circ$ with respect to the Sun--Galactic Center line. The Sun is placed at a Galactocentric distance of $8.1$ kpc. To visualize the geometric coverage, we plot four thin dashed lines representing lines of sight at Galactic longitudes $\ell= \pm 10^\circ$ and $\pm 20^\circ$, which bracket the region of interest for the Tremaine-Weinberg application. The thick black dashed line delineates the major axis and approximate length of the stellar bar. Panel (b) shows the recovered $\Omega_{\rm p}$ as a function of longitude. In the range $|\ell| \approx 5^\circ$, the method successfully recovers a value consistent with the true pattern speed (red horizontal line). However, the profile exhibits sharp divergences at specific longitudes (e.g., near $\ell=0^\circ$ and $\ell \approx 12^\circ$). These are not physical variations but mathematical artifacts. They occur where the integral in the denominator—the density-weighted moment of the transverse position—approaches zero due to the specific symmetry of the density distribution along that line of sight relative to the observer.

To overcome these local instabilities, we examine the correlation between the numerator and denominator terms directly, as shown in Figure~\ref{fig:mw_validation} panel (c). By plotting $Num_i$ against $Denom_i$ for all angular bins, we observe a tight linear relation. The slope of this relation provides the most robust estimate of the standard pattern speed, yielding $\Omega_{\rm p} = 45.1 \pm 2.0 \, \mathrm{km \, s^{-1} \, kpc^{-1}}$, in excellent agreement with the simulation ground truth.


\bibliography{sample701}{}
\bibliographystyle{aasjournalv7}



\end{document}